\documentclass{opticajnl}
\journal{opticajournal} 

\setboolean{shortarticle}{false}

\usepackage{lineno}

\DeclareUnicodeCharacter{3B1}{\ensuremath{\alpha}}

\title{Optical Footprint of Ghost and Leaky Hyperbolic Polaritons}

\author[1, *]{Mark Cunningham}
\author[1]{Adam L. Lafferty}
\author[2]{Mario González-Jiménez}
\author[1]{Rair Macêdo}

\affil[1]{James Watt School of Engineering, 
Electronics \& Nanoscale Engineering Division, 
University of Glasgow, Glasgow G12 8QQ, United Kingdom}
\affil[2]{School of Chemistry,
University of Glasgow, Glasgow G12 8QQ, United Kingdom}
\affil[*]{m.cunningham.2@research.gla.ac.uk}

\begin{abstract}
Manipulating hyperbolic polaritons at infrared frequencies has recently garnered interest as it promises to deliver new functionality for next-generation optical and photonic devices.
This study investigates the impact of the crystal's anisotropy orientation on the Attenuated Total Reflection (ATR) spectra, more specifically, revealing the optical footprint of elliptical, ghost (GHP) and leaky (LHP) hyperbolic polaritons. 
Our findings reveal that the ATR spectra of GHPs exhibit a distinct hyperbolic behaviour which is similar to that recently observed using s-SNOM techniques. 
Similarly, the ATR spectra of LHPs show its clear lenticular behaviour; however, here we are able to discern the effects of large asymmetry due to cross-polarisation conversion when the crystal anisotropy is tilted away from the surface.
Furthermore, we demonstrate that by controlling the anisotropy orientation of hyperbolic media it is possible to significantly alter the optical response of these polaritons. 
Thus, our results provide a foundation for the design of direction-dependent optical devices.
\end{abstract}

\setboolean{displaycopyright}{false} 

\begin{document}

\maketitle

\section{Introduction}

Hyperbolic materials have attracted 
significant 
interest for controlling light at the nanoscale \cite{teng_steering_2024, zhang_ultrafast_2023, hu_real-space_2023}, 
due to their intrinsic hyperbolic dispersion and strong electromagnetic field confinement \cite{dumelow_chapter_2016}. 
They have been shown to support surface polaritons \cite{macedo_oriented_2018, voronin_fundamentals_2024, zhang_unique_2023, ma_ghost_2021, sun_volume-confined_2022, zhang_hybridized_2021, ni_observation_2023, dai_manipulation_2018, schuller_surface_1975}, 
with long-range and low-loss propagation of subwavelength information \cite{ma_ghost_2021, ni_observation_2023}, 
guided waves \cite{ji_cladding-free_2024, heydari_highly-controllable_2024, fang_hyperbolic_2024, f_tresguerres-mata_observation_2024}, 
and negative refraction \cite{macedo_oriented_2018, zhang_negative_2019, moradi_reflection_2023, wu_optical_2021, dumelow_chapter_2016}.
In these materials, hyperbolic dispersion results from the principal elements of the permeability or permittivity tensor possessing opposite signs \cite{fang_hyperbolic_2024, macedo_oriented_2018, dumelow_chapter_2016, ma_ghost_2021, ni_observation_2023, matson_controlling_2023, jones_controlling_2023, passler_hyperbolic_2022, mock_effective_2024}.
This is typically a consequence of anisotropy due to elementary resonances in matter and can emerge across a vast range of frequencies depending on the mechanism (e.g. phonon resonances in the infrared \cite{estevam_da_silva_far-infrared_2012, you_strong_2024, heydari_highly-controllable_2024, chen_active_2024} or magnetic resonances \cite{macedo_engineering_2019, wang_surface_2023, dantas_surface_2023, costa_strongly_2023, bludov_hybrid_2019} in the GHz and THz frequency range).
The role of anisotropy orientation with respect to the interface has recently received attention for creating direction-dependent optical devices \cite{lekner_reflection_1991, wu_asymmetric_2022, jones_controlling_2023, macedo_oriented_2018, wu_emergent_2019, passler_layer-resolved_2020,passler_layer-resolved_2023, hu_topological_2020, jia_two-dimensional_2022, low_polaritons_2017, duan_twisted_2020, capote-robayna_twisted_2022, zhao_ultralow-loss_2022, caldwell_atomic-scale_2016} and has, more recently, led to the discovery of peculiar entities including Ghost Hyperbolic Polaritons (GHPs) \cite{ma_ghost_2021} and Leaky Hyperbolic Polaritons (LHPs) \cite{ni_observation_2023}.

Originally observed in calcite \cite{ma_ghost_2021}, a natural hyperbolic material, GHPs have been characterised as virtual surface phonon-polariton modes \cite{hao_ghost_2023, wang_ghost_2023}, devoid of an electrostatic limit (with damped propagation even in the absence of material loss \cite{venturi_visible-frequency_2024}) with characteristic high-resolution, ray-like propagation at the material surface over distances up to $20~\mu$m \cite{ma_ghost_2021}.
These are a consequence of the anisotropy axis being neither parallel nor perpendicular to the sample surface; this, in turn, gives rise to phase wavefronts slanted away from the surface of the crystal, Poynting Vectors parallel to the surface, and exponential decay away from the surface. 
Since their emergence, spatiotemporal ultrafast dynamics of GHP nanolight pulse propagation has also been studied \cite{zhang_ultrafast_2023} as well as planar junctions and anisotropic metasurfaces that can support 'ghost line waves' that propagate unattenuated along the line interface, with phase oscillations combined with evanescent decay away from the interface \cite{moccia_ghost_2023}.
In the visible light range, ghost modes have been observed in the bulk plasmonic material MoOCL\textsubscript{1} \cite{venturi_visible-frequency_2024}. 
Remarkably though, they do not require an interface or a tilted anisotropy axis away from the surface.

LHPs, on the other hand, are hybridisations of propagating ordinary bulk modes and extraordinary surface waves \cite{ni_observation_2023}. 
They originate from the physics of leaky waves, studied since the 1940s, which are guided modes that propagate along open wave-guiding structures \cite{goldstone_leaky-wave_1959, oliner_guided_1959}.
Unlike conventional surface waves, leaky waves possess complex propagation constants, allowing energy radiation into free space as they travel \cite{tamir_leaky_1975,monticone_leaky-wave_2015, tamir_lateral_1971}.
Leaky wave antennas have utilised plasmonic metamaterials with Epsilon-Near-Zero (ENZ) dielectric constants to enhance directionality \cite{alu_epsilon-near-zero_2007}. 
Anisotropic ENZ materials with tunable transverse dielectric responses and low magnetic permeability materials with strong tangential magnetic fields have also been employed to control emission and directivity \cite{halterman_controlled_2011}.
Hyperbolic materials are ideal candidates for supporting leaky waves, particularly around longitudinal optic phonon frequencies where dielectric tensor components are small \cite{dumelow_chapter_2016}. These materials offer advantages over plasma layers, which only support leaky waves above their plasma frequency.
Additionally, LHPs have been shown to exhibit tilted wavefronts into the bulk and Poynting vectors canted away from the interface \cite{ni_observation_2023}.

These novel polariton modes have been, very recently, studied in-depth using s-SNOM and further supported by spectroscopy. 
Here, we demonstrate how GHPs and LHPs manifest through attenuated total reflection (ATR) spectroscopy in bulk hyperbolic crystals, specifically investigating their direction-dependent optical footprint across the crystal's surface. 
By investigating how these polaritons appear in the far-field using more widely-available infrared spectroscopy, we use ATR to investigate the conditions under which LHPs and GHPs may be supported in bulk hyperbolic media; using crystal quartz as the example material. 
This should provide valuable insight for integrating these materials into, as well as exploiting these properties for, novel optical devices.

\begin{figure*}[!ht]
  \centering  
      \includegraphics[width=0.5\linewidth, keepaspectratio]{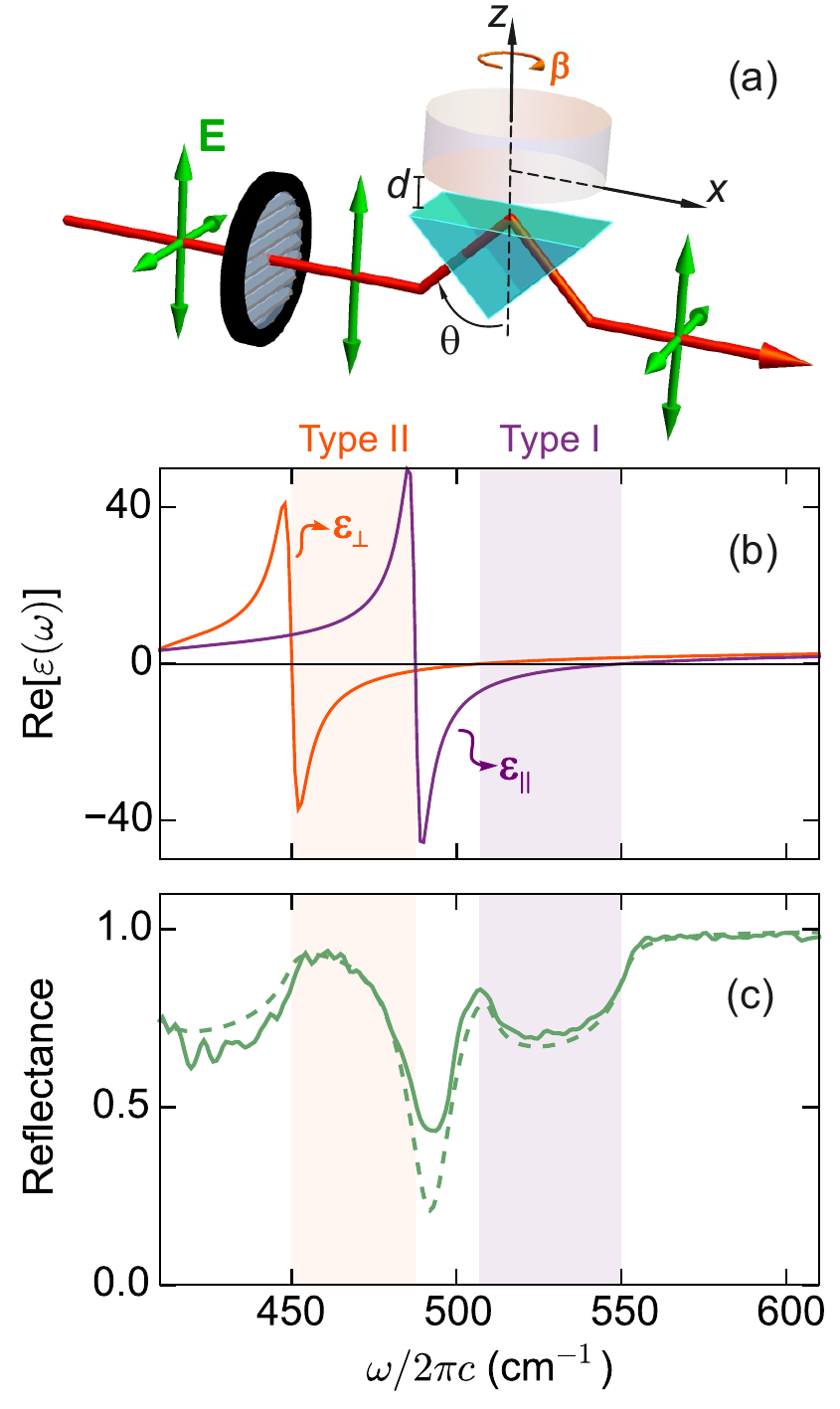}
  \caption{(a) Experimental setup geometry, where a polariser is used to generate a TM-polarised beam which is incident at the surface of crystal quartz at an incident angle of $\theta=$~45$^\circ$ from a dielectric prism ($\varepsilon_p=$~5.5). 
  (b) Real part of the principal components of the dielectric function of quartz in the
    frequency range 410~cm\textsuperscript{-1} to 610~cm\textsuperscript{-1}.
  (c) Theoretical (dashed line) and experimental (solid line) reflectance spectra [using the set up shown in (a)] for a crystal quartz sample whose anisotropy lies along the $z$ axis and an air gap of $d=$~2.0~$\mu$m.}
  \label{fig:IntroDiagrams}
\end{figure*}

We employ ATR in the Otto configuration \cite{falge_dispersion_1973, anderson_attenuated_2013, passler_layer-resolved_2023}, as shown in Fig.~\ref{fig:IntroDiagrams}(a), where reflection occurs at the boundary between a prism and the crystal supporting surface polaritons. In turn, the prism with high dielectric constant, $\varepsilon_p$, generates a high in-plane momentum with wavevector $k_x = k_0 \sqrt{\varepsilon_p} \sin{\theta}$, where $k_0 = \omega/c$ and $\theta$ is the incident angle of the infrared waves with angular frequency $\omega$ that couple to polariton modes. 
Note, that if an air gap $d$ is included between the prism and the crystal and radiation is incident at an angle higher than the critical angle of the prism/air interface, total internal reflection occurs and only evanescent modes travel across the air gap.

The natural hyperbolic material used here, crystal Quartz, supports hyperbolic polaritons due to infrared-active phonon resonances. 
This makes the crystal anisotropic with a diagonal dielectric tensor.
For instance, in the frequency range between 410 and 610~cm\textsuperscript{-1}, corresponding to free-space wavelengths between 16 and 25~$\mu$m, there are two regions where the principal components of the dielectric tensor possess opposite signs, as shown in Fig.~\ref{fig:IntroDiagrams}(b). 
Therefore, hyperbolic behaviour is observed: a Type I region, i.e. the extraordinary component $\varepsilon_\parallel$ of the dielectric tensor is negative while the ordinary $\varepsilon_\perp$ is positive, between 510 and 550~cm\textsuperscript{-1}, and a Type II region, $\varepsilon_\parallel>0$ and $\varepsilon_\perp<0$, between between 450 and 480~cm\textsuperscript{-1}.

To build context, let us look at the case where $\varepsilon_{\parallel}$ is perpendicular to the surface of the crystal (along $z$) and the crystal axes are aligned with the laboratory axes. 
SPhPs typically propagate along the crystal's surface with an exponential decay of amplitude away from the surface. 
In the case shown in Figure 1(a), SPhPs couple with TM-polarised infrared radiation.
Assuming propagation along $x$ and decay along $z$, given by the parameters $\alpha_0$ (air) and $\alpha$ (hyperbolic material), plane wave solutions take the following form in air:

\begin{equation}
\mathbf{H}(x,z,t) = \hat{y}H e^{i(k_x x - \omega t)} e^{\alpha_0 z}
\label{eqn:AirWaveNEW}
\end{equation}

\noindent and in the hyperbolic material:

\begin{equation}
\mathbf{H}(x,z,t) = \hat{y} H e^{i(k_x x - \omega t)} e^{-\alpha z}
\label{eqn:QuartzWaveNEW}
\end{equation}
Using Maxwell's equations, similarly to \cite{macedo_engineering_2019}, and matching its solutions inside and outside the hyperbolic material, the boundary conditions allow us to find the following dispersion relation for the surface polaritons:

\begin{equation}
0 = \alpha_0 + \frac{\alpha}{\epsilon_{xx}}.
\label{eqn:DispersionRelationEquationNEW}
\end{equation}
Since $\alpha_0$ and $\alpha$ are positive (to ensure decay away from the surface) and the anisotropy axis is oriented along $z$ such that $\varepsilon_{xx}$ = $\varepsilon_{\perp}$, ~\ref{eqn:DispersionRelationEquationNEW} can only be satisfied when $\varepsilon_{\perp}<0$. The corresponding ATR response is shown in Fig.~\ref{fig:IntroDiagrams}(c) where a distinct minimum in the reflection is observed between hyperbolic regions, corresponding to where the condition set out above is met. 
In this case, evanescent waves in air, generated by the total internal reflection in the prism, couple to SPhPs in the hyperbolic material.
Here, we used a coupling prism with dielectric constant $\varepsilon_p =$~5.5 and an air gap of $d =$~1.5~$\mu$m at a fixed incident angle of $\theta=$~45$^{\circ}$, corresponding to an in-plane momentum of $k_x/k_0 =$~1.66 to induce total internal reflection.

\begin{figure}[!ht]
\centering
\includegraphics[width = 0.5\linewidth, keepaspectratio]{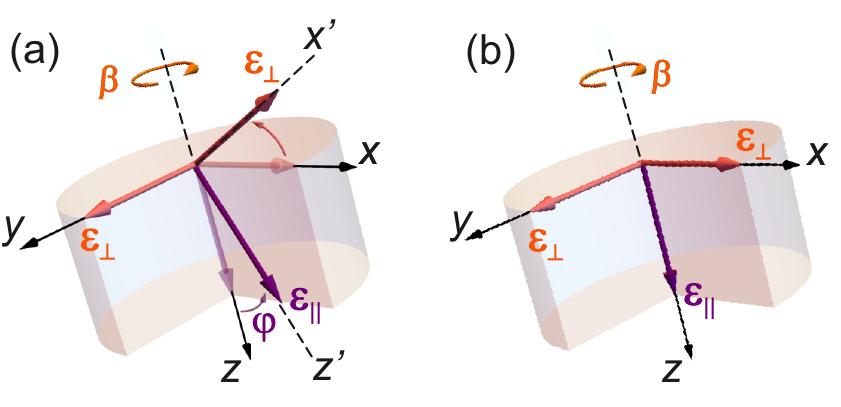}
\caption{a) Geometry of the dielectric components with respect to the laboratory axis when introducing the angle $\varphi$, b) Geometry of the dielectric components with respect to the laboratory axis when introducing the angle $\beta$.}
\label{fig:BetaATRDiagrams}
\end{figure}

While the behaviour of SPhPs detailed above is well known, it can be drastically modified by controlling the orientation of the anisotropy.
One way to achieve this is by tilting the anisotropy axis by an angle $\varphi$ away from the $z$ axis, as shown in Fig.~\ref{fig:BetaATRDiagrams}(a), so that $\varepsilon_\parallel$ is neither perpendicular nor parallel to the interface. 
When discussing this angle, a few special configurations are worth noting, namely $\varphi=$~0$^{\circ}$ and $\varphi=$~90$^{\circ}$ which correspond to $\varepsilon_{\parallel}$ aligning perpendicular and parallel to the surface, respectively.
Altering $\varphi$ will effectively rotate the hyperbolic dispersion \cite{macedo_oriented_2018}, introducing off-diagonal components to the dielectric tensor \cite{macedo_oriented_2018, wu_asymmetric_2022}.
Another way to modify the behaviour of SPhPs is by introducing an azimuthal rotation around the $z$ axis by an angle $\beta$ with respect to the incidence plane (here taken as $x$-$z$) as shown in Fig.~\ref{fig:BetaATRDiagrams}(b), which mimics the effect of an in-plane wavevector $k_y$. 
Therefore, a special case includes $\varphi=$~90$^{\circ}$ and $\beta = 90^{\circ}$ making $\varepsilon_{\perp}$ and $\varepsilon_{\parallel}$ aligned along the $x$ and $y$ axes, respectively. 
Another special case is when $\varphi=$~0$^{\circ}$. In this case, the components on the dielectric tensor aligned with $x$ and $y$ are both equal to $\varepsilon_\perp$, as shown in Fig.~\ref{fig:BetaATRDiagrams}(b). Therefore, there is no azimuthal dependence of the spectrum shown in Fig.~\ref{fig:IntroDiagrams}(c).

\begin{figure*}[!ht]
\centering
\includegraphics[width=\linewidth]{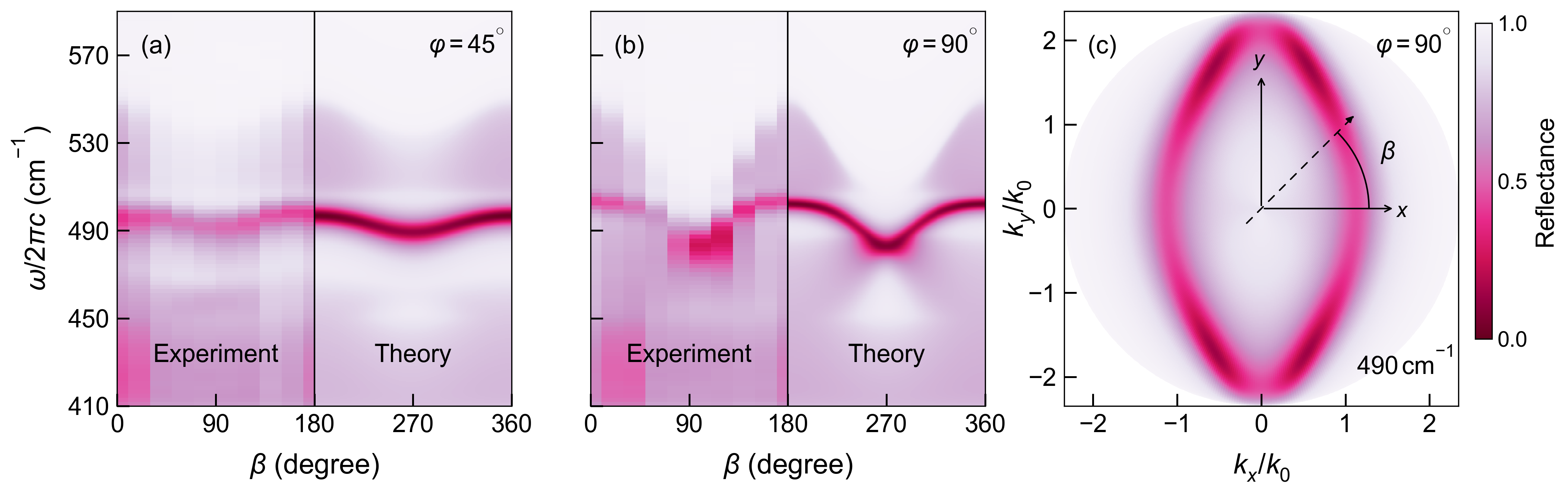}
\caption{The dependence of ATR in crystal quartz on the azimuthal angle $\beta$.
a)$\varphi=$~45$^{\circ}$,
b) $\varphi=$~90$^{\circ}$, with $\frac{k_x}{k_0} = 1.66$. $\epsilon_p =$~5.5 and $d =$~2~$\mu$m.
c) ATR of Quartz at $\omega/2\pi c=$~490~cm\textsuperscript{-1} when $\varphi=$~90$^{\circ}$, with the radius corresponding to $k_x$ where $\epsilon_p =$~5.5, and the azimuthal angle corresponding to the angle $\beta$, with $d =$~2~$\mu$m}
\label{fig:ellipticalsurfaceplots}
\end{figure*}

On the other hand, an azimuthal rotation will introduce significant change to this spectrum if $\varphi\neq$~0 as shown in in Fig.~\ref{fig:ellipticalsurfaceplots} where both theoretical and experimental results for $\varphi=$~45$^{\circ}$ [Fig.~\ref{fig:ellipticalsurfaceplots}(a)] and $\varphi=$~90$^{\circ}$ [Fig.~\ref{fig:ellipticalsurfaceplots}(b)] are given. 
Clear agreement is shown between theory and experiment and, in both cases, we can see a sharp dip in reflectivity between the two hyperbolic regions, corresponding to the existence of a surface polariton. 
When $\varphi=$~45$^{\circ}$ as shown in Fig.~\ref{fig:ellipticalsurfaceplots}(a), the surface wave displays a sinusoidal azimuthal dependence. The excitation frequency reaches its maximum value of $\omega /2 \pi c =$~500~cm\textsuperscript{-1} at azimuthal angles $\beta =$~0$^{\circ}$ and 180$^{\circ}$, while the minimum value of $\omega /2 \pi c =$~480~cm\textsuperscript{-1} occurs at $\beta =$~90$^{\circ}$ and 270$^{\circ}$. 
We note that the surface wave exists entirely within a region where bulk propagation is not allowed, shown by the rectangular shape of high reflectivity around the surface wave.

When we rotate the anisotropy further ($\varphi=$~90$^{\circ}$) , as shown in Fig.~\ref{fig:ellipticalsurfaceplots}(b), the dependence on azimuthal orientation now takes a rectified sinusoidal shape. 
This frequency reaches its maximum value of $\omega /2 \pi c = $~505~cm\textsuperscript{-1} at azimuthal angles $\beta = 0^{\circ}$ and 180$^{\circ}$, while the minimum value of 475~cm\textsuperscript{-1} occurs at $\beta =$~90$^{\circ}$ and 270$^{\circ}$.

So far, we have focused on the azimuthal dependence of the ATR spectra at a fixed $k_x/k_0$ value for a range of frequencies, we will now turn to the case of a fixed frequency between the two hyperbolic regions where the surface polariton exists. This is shown in Fig.~\ref{fig:ellipticalsurfaceplots}(c) where the behaviour of varying $k_x/k_0$ is given for $\varphi=$~90$^{\circ}$ and at 490~cm\textsuperscript{-1}.
We can see that the surface polariton displays an elliptical profile with respect to the azimuthal angle $\beta$ , which is to be expected where all dielectric tensor components are negative \cite{ni_long-lived_2021, passler_hyperbolic_2022}. 
We can see that a smaller $k_x/k_0$ is needed when $\beta = 0^{\circ}$, and a larger value is needed towards $\beta = 90^{\circ}$. This is because $\epsilon_{\parallel}$ has a much larger negative value than $\epsilon_{\perp}$, so less in-plane momentum is required to couple with the surface polariton when the anisotropy is aligned with the $x$ axis.

Now that we have shown how the orientation of the anisotropy axis can greatly impact the existence of surface polaritons in crystal quartz, we can redirect our analysis towards the reflective traits of GHPs and LHPs.

\section{Ghost Polaritons}

We have seen how surface waves behave in anisotropic materials, namely their elliptical behaviour with respect to $\beta$ (or in-plane wave-vector). 
Now, let us change the frequency region of interest; instead of looking at regions where no propagation is allowed, thus surface waves can emerge, we turn to hyperbolic regions where propagation is allowed; beginning with the Type II hyperbolic region.
In Type II hyperbolic regions there are two main cases to consider: (a) the anisotropy perpendicular to the crystal surface (here along $z$), and (b) when the anisotropy aligns with the crystal's surface.
The first case would be similar to the classic surface polariton which we have discussed in detail as both components of the dielectric tensor along the surface are negative and despite $\varepsilon_{zz}$ being positive, the condition in Eq. 3 could still be met for any $\beta$. 
The second case, on the other hand, is far more complex. 
For instance, when $\varphi=$~90$^\circ$ and $\beta=$~0$^\circ$ ($\varepsilon_{\parallel}$ lies along $x$ and $\varepsilon_\perp$ is along $y$).
Eq.~3 can thus only be satisfied if $\beta =$~90$^{\circ}$, resulting in $\varepsilon_{\parallel}$ aligning with $y$ and $\varepsilon_\perp$ along $x$. 
In this orientation, and considering the incidence plane to be the $x-z$ plane, crystal quartz behaves entirely like a metal ($\varepsilon_{xx} = \varepsilon_{zz} = \varepsilon_{\perp}$) and thus, supports a surface polariton.

\begin{figure*}[!ht]
\centering
\includegraphics[width=\linewidth]{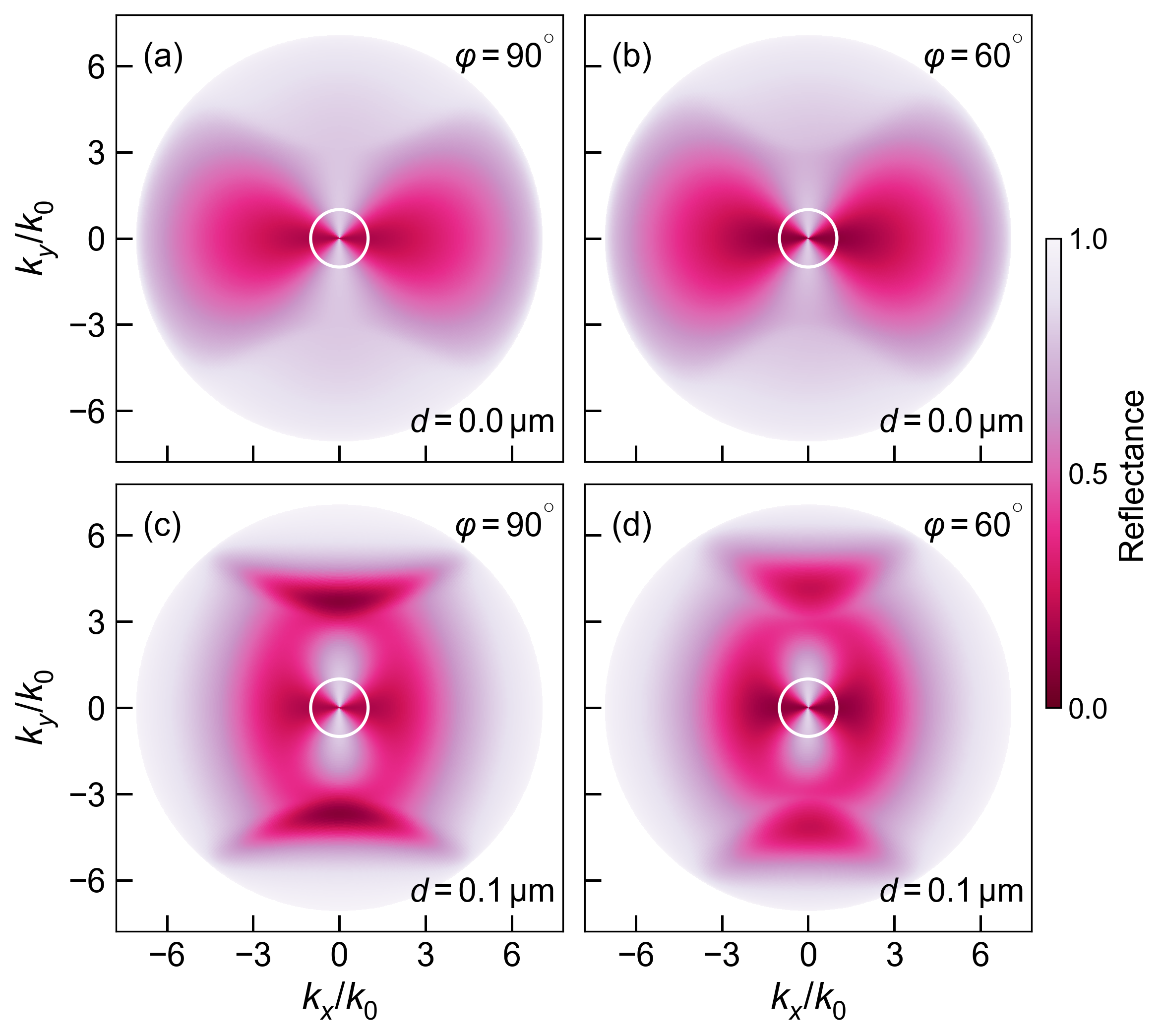}
\caption{
ATR spectra for crystal quartz at $\omega/2\pi c=$~460~cm\textsuperscript{-1}, with the radius corresponding to $k_x$ where $\epsilon_p =$~50, and the azimuthal angle corresponding to the angle $\beta$. 
The white circle denotes where $k_x/k_0=$~1.
In a) there is no air gap ($d =$~0~$\mu$m) and$\varphi=$~90$^{\circ}$.
In b) the anisotropy orientation is unchanged ($\varphi=$~0$^{\circ}$) and an air gap is introduced to study the GHP, with $d =$~0.1~$\mu$m.
In c), the air gap is removed ($d =$~0~$\mu$m) and the anisotropy is rotated to $\varphi=$~60$^{\circ}$ to alter the hyperbolic dispersion.
In d), the anisotropy orientation is unchanged ($\varphi=$~60$^{\circ}$) and an air gap is introduced again to study the GHP, with $d =$~0.1~$\mu$m.
}
\label{fig:ghostpolarplots}
\end{figure*}

To illustrate this we calculate the ATR behaviour at $\omega/2\pi c=$~460~cm\textsuperscript{-1}, shown in Fig.~\ref{fig:ghostpolarplots}.
First we look at the case where $\varphi=$~90$ ^{\circ}$ but no air gap ($d=$~0~$\mu$m). 
This is shown in Fig.~\ref{fig:ghostpolarplots}(a), where we can see the clear bulk hyperbolic dispersion traced by propagating radiation (i.e. continuous dips in reflectance). 
Propagation is mostly supported when $\beta=$ ~0$^{\circ}$ or 180$^\circ$, reducing in intensity for higher $k_x$ values. 
On the other hand, propagation is completely forbidden when $\beta=$~90$ ^{\circ}$ due to the quartz behaving entirely like a metal.
In Fig.~\ref{fig:ghostpolarplots}(b), we show the ATR when an air gap of $d=$~0.1~$\mu$m is present so that evanescent waves can couple to surface waves. We observe a hyperbola-shaped drop in reflectivity, indicative of a surface wave, in the region where bulk propagation is forbidden (i.e. upper region of the plot). 
As expected, it requires smaller $k_x$ values where $\beta=$~90$ ^{\circ}$, requiring more in-plane momentum when azimuthal rotation is introduced, until the surface wave is no longer supported at $\beta= 45 ^{\circ}$ and $\beta= 135 ^{\circ}$ where the bulk hyperbolic band exists. 
This highly angular dispersion mirrors that found in the original s-SNOM results of the original GHP discovery \cite{ma_ghost_2021}.

Earlier, we have also seen how combining a tilted anisotropy axis (with respect to the surface and a combination) with in-plane wave-vectors can yield an extra degree of control of the surface polariton behaviour. 
Since tilting the anisotropy in such a way has been used as a route to introduced GHPs \cite{ma_ghost_2021}, we now look at the effect of tilting the anisotropy away from the surface with $\varphi=$~60$^{\circ}$, shown in Fig.~\ref{fig:ghostpolarplots}(c)-(d), again at $\omega/2\pi c=$~460~cm\textsuperscript{-1}.
For the first case, Fig.~\ref{fig:ghostpolarplots}(c) we again look at no air gap ($d=$~0~$\mu$m). 
The propagating behaviour tracing the hyperbolic dispersion is very similar to Fig.~\ref{fig:ghostpolarplots}(a), except the reflectivity dip occurs at a slightly larger range of azimuthal angles.
Introducing an air gap of $d=$~0.1~$\mu$m, shown in Fig.~\ref{fig:ghostpolarplots}(d), clearly displays the impact of rotating the anisotropy in the form of a GHP. 
Qualitatively, the shape of the ATR spectra is very similar to the untilted case. 
However, the surface wave component is now supported by a narrower range of azimuthal angles, centred on $\beta= 90 ^{\circ}$. 
The drop in reflectivity is less intense, showing the GHP moving away from the bulk hyperbolic dispersion.
Moreover, tilting the anisotropy away from the surface means that the positive extraordinary dielectric tensor component will now provide propagation characteristics to the polariton (instead of a 'true' surface polariton). This has been previously explained as tilted wavefronts into the bulk of the material and as long-distance propagation along the surface \cite{ma_ghost_2021}. 
This anisotropy tilting introduces more pronounced optical effects associated with large anisotropy such as asymmetric cross-polarisation conversion \cite{wu_asymmetric_2022}, where details are shown in Appendix D.

\begin{figure}[!ht]
\centering
\includegraphics[width = \linewidth, keepaspectratio]{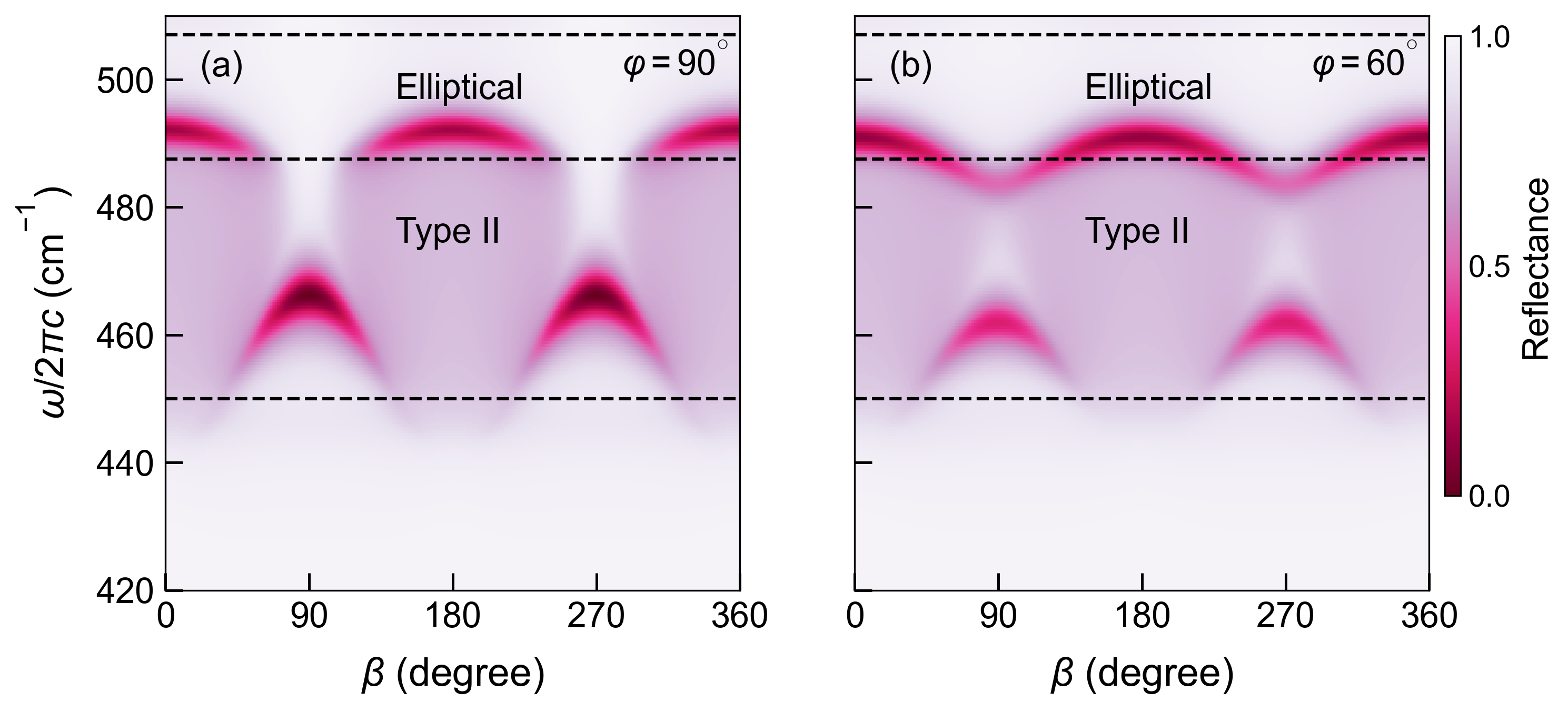}
\caption{
The dependence of ATR in Quartz on the azimuthal angle $\beta$, with $\frac{k_x}{k_0} = 5$ and $\epsilon_p =$~50. An air gap is included ($d =$~0.1~$\mu$m) to probe the GHP.
In a), the anisotropy is aligned with the interface ($\varphi=$~90$^{\circ}$). In b), the anisotropy is rotated below the interface where $\varphi=$~60$^{\circ}$.
}

\label{fig:ghostazimuthalplots}
\end{figure}

In order to obtain the results above we needed to modify a few parameters, namely increase the prism dielectric constant compared to the experimental case shown in earlier sections. 
This is necessary to be able to generate high enough in-plane momentum to probe a large portion of the hyperbolic dispersion.
As a consequence of the larger in-plane momentum, we also needed to reduce the air gap thickness to support critical coupling (see Appendix E for details on the effect of varying $d$).
While we illustrate this behaviour at a single frequency, in order to address behaviour observed in S-NOM, the frequency dependent behaviour is also vastly different to what we previously shown for true surface polaritons. For more information on the ATR response at specific frequencies, see Appendix F.
In Fig.~\ref{fig:ghostazimuthalplots} we show the ATR spectra over the frequency range 430-500~cm\textsuperscript{-1} at a fixed $k_x/k_0=$~5. 
For $\varphi = $~90$^{\circ}$, in Fig.~\ref{fig:ghostazimuthalplots}(a), the elliptical surface polariton examined previously splits into segments at ~490~cm\textsuperscript{-1}, positioned above the bulk propagation linked with Type~II hyperbolic dispersion. 
A distinct GHP emerges between 450~-~470~cm\textsuperscript{-1}, peaking around $\beta=$~90$^{\circ}$, 270~$^{\circ}$, where $\epsilon_{\perp}$ aligns on both $x$ and $z$ axes.
The GHP's frequency-$\beta$ relationship resembles an upward arrow, segregating bulk-propagation zones in the Type II hyperbolic region.
At $\varphi=$~60$^{\circ}$, shown in Fig.~\ref{fig:ghostazimuthalplots}(b), the GHP's intensity reduces. 
The elliptical surface wave, rather than being separated, now connects via a slight reflectivity dip at 485~cm\textsuperscript{-1} near $\beta=$~90$^{\circ}$, 270~$^{\circ}$, also increasing the range of $\beta$ angles for which there is propagation within the Type II region. 
Increasing the range of angles for which there is propagation inside the Type II regions seems to also alter the GHP, decreasing its reflectance intensity, consistent with true surface waves being more pronounced where bulk propagation is absent \cite{macedo_oriented_2018}.

\section{Leaky Polaritons}

\begin{figure}[!ht]
\centering
\includegraphics[width = \linewidth, keepaspectratio]{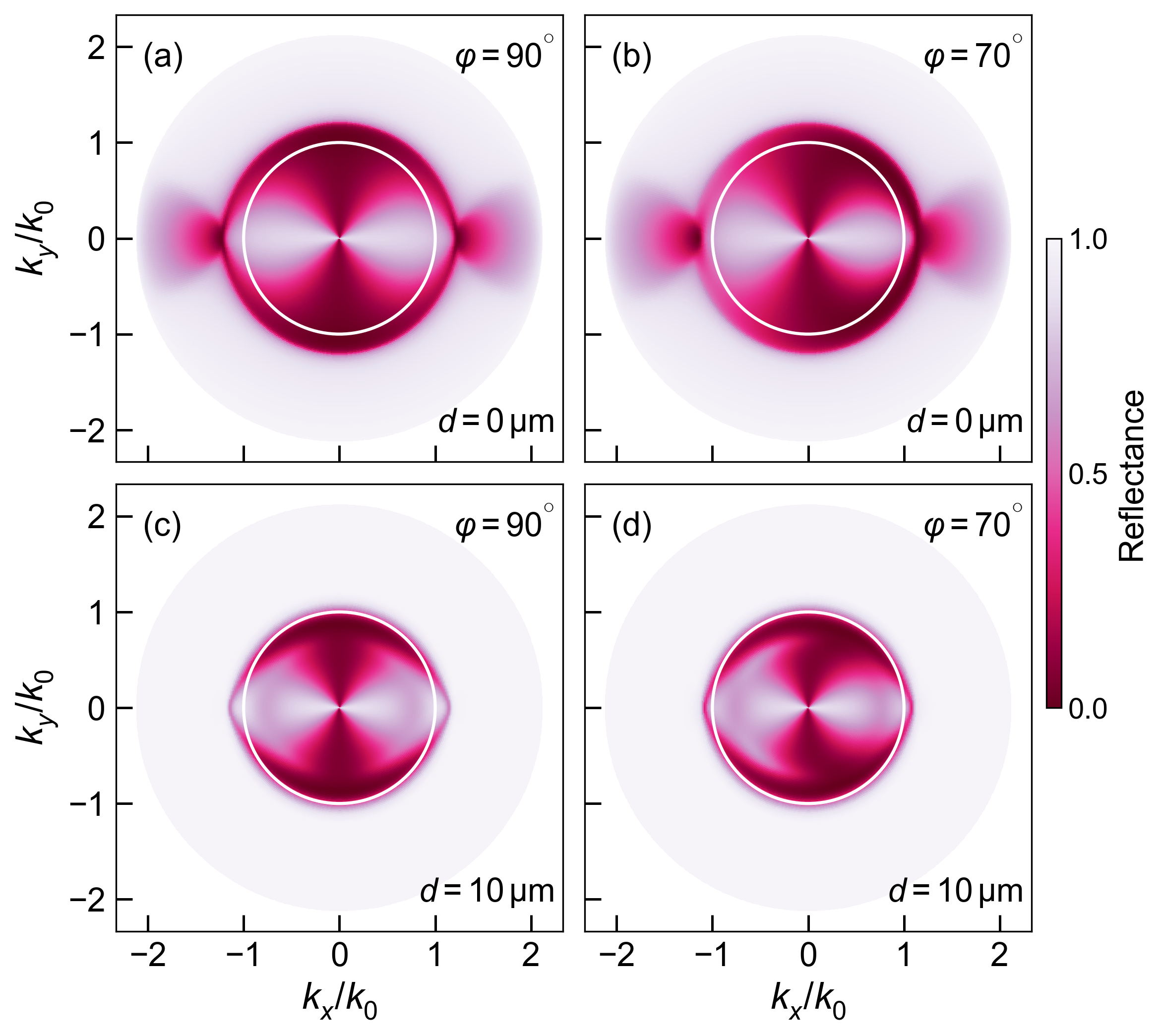}
\caption{
ATR spectra for crystal quartz at $\omega/2\pi c=$~545~cm\textsuperscript{-1}, with the radius corresponding to $k_x$ where $\epsilon_p =$~2.2, and the azimuthal angle corresponding to the angle $\beta$. 
In a) there is no air gap ($d =$~0~$\mu$m) and $\varphi=$~90$^{\circ}$.
In b) the anisotropy orientation is unchanged ($\varphi=$~90$^{\circ}$) and an air gap is introduced to study the LHP, with $d =$~10~$\mu$m.
In c), the air gap is removed ($d =$~0~$\mu$m) and the anisotropy is rotated to $\varphi=$~70$^{\circ}$ to alter the hyperbolic dispersion.
In d), the anisotropy orientation is unchanged ($\varphi=$~70$^{\circ}$) and an air gap is introduced again to study the LHP, with $d =$~10~$\mu$m.
The white, solid lines denote the light line (i.e. $k_x/k_0 =k_y/k_0=$~1.0).
}
\label{fig:leakypolarplots}
\end{figure}

Similarly to GHPs, leaky polaritons have recently been studied using s-SNOM where the direction of the anistropy with respect to the crystal surface has been shown to play a crucial role \cite{ni_observation_2023}.
To understand how LHPs manifest in ATR spectra, we now shift our focus to Type I hyperbolic regions where these waves have been shown to emerge in calcite \cite{ni_observation_2023}. 
To allow radiation leakage in an ATR setup, an air gap must be introduced to include the coupling of evanescent waves with the leaky wave.
Leaky waves experience an exponential increase in air \cite{monticone_leaky-wave_2015}, which in the past has led them to be known as 'mathematically improper' \cite{zheng_leaky-wave_2023}, meaning that quite a large air gap (as used here) will be necessary to allow coupling. 
As leaky waves appear at low wavevectors, we only observe surface wave-like to be supported slightly beyond the critical angle, close to $k_x/k_0=$~1.
Fig.~\ref{fig:leakypolarplots} shows the ATR behaviour of leaky polaritons at 545~cm\textsuperscript{-1} in crystal quartz.

In Fig.~\ref{fig:leakypolarplots}(a), where $\varphi=$~90$^{\circ}$ and no air gap, we can see that the ATR spectra traces the hyperbolic dispersion (left- and right-hand side) due to the in-plane anisotropy, with two extraordinary cones on either side of the bulk ordinary cone. 
Outside of the ordinary cone, propagation is mostly supported when $\beta=$ ~0$^{\circ}$ or ~180$^{\circ}$, reducing in intensity for higher $k_x$ values.
In Fig.~\ref{fig:leakypolarplots}(b), we introduce an air gap $d=$~10~$\mu$m which induces the emergence of leaky polaritons. 
We observe a lenticular-shaped drop in reflectivity, indicative of the leaky polariton \cite{ni_observation_2023}. 
What is interesting is that this drop in reflectivity is only at a small $k_x$ interval larger than the critical angle, but it is within the bulk ordinary circle bounds of Fig.~\ref{fig:leakypolarplots}(a). 
The leaky polariton is supported mostly when $\beta= 0 ^{\circ}$, where metallic contribution from the negative dielectric tensor component is at its maximum. 
At $\beta= 90 ^{\circ}$, only non-metallic (positive) dielectric tensor components contribute, therefore no surface-like wave phenomena are supported.

The effect of tilting the anisotropy away from the surface ($\varphi=$~70$^{\circ}$) is shown in Fig.~\ref{fig:leakypolarplots}(c)-(d) for both no air gap and $d=$~10~$\mu$m. 
When probing bulk waves (no air gap) we can see that rotating the anisotropy causes the extraordinary cones to move inward, overlapping with the bulk ordinary cone, facilitating radiation leakage \cite{ni_observation_2023}. Notable asymmetry is observed due to cross-polarisation conversion, which has been studied previously, albeit not using ATR \cite{wu_asymmetric_2022}.
On the other hand, introducing an air gap causes the lenticular shape of the leaky polariton to move inwards too. 
The reflectivity drop is still shown slightly beyond the critical angle, so that evanescent coupling is still occuring, but within the bulk ordinary cone. 
We also note the stark asymmetry within the critical angle due to much larger cross polarisation conversion than what was observed with no air gap (see Appendix G for details on cross polarisation conversion due to LHPs).

\begin{figure}[!ht]
\centering
\includegraphics[width = \linewidth, keepaspectratio]{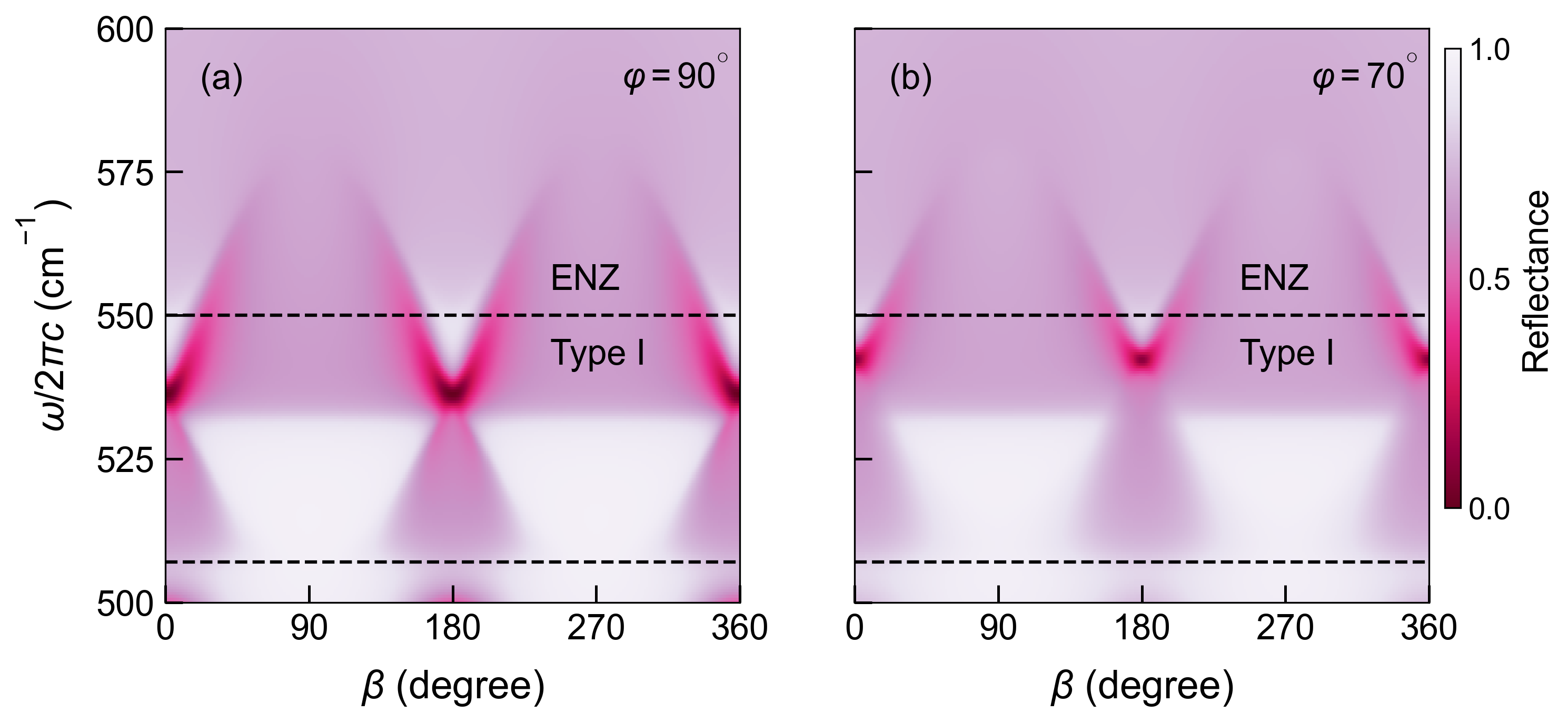}
\caption{
The dependence of ATR in Quartz on the azimuthal angle $\beta$, with $k_x/k_0=$~1.05~ and $\epsilon_p =$~5.5. An air gap is included ($d =$~10~$\mu$m) to probe the LHP.
In a), the anisotropy is aligned with the interface ($\varphi=$~90$^{\circ}$). In b), the anisotropy is rotated below the interface such that $\varphi=$~70$^{\circ}$.
}

\label{fig:leakyazimuthalplots}
\end{figure}

In Fig.~\ref{fig:leakyazimuthalplots}(a), we see the azimuthal dependence leaky polaritons in crystal quartz when the anisotropy is aligned with the surface, and an incident angle of $\theta=$~XX$^\circ$ (corresponding to $k_x/k_0=$ ~1.05 for a prism of $\varepsilon_p=$~5.5). 
The reflectivity dip takes a downward-arrow shape, centred at azimuthal angles of $\beta=$~0 and ~180$^\circ$ which correspond to the anisotropy aligning with the $x$ axis and separating into two branches that widen as the frequency increases.
This behaviour is directly linked to the isofrequency contours of dispersion with lenticular shape observed in near-field imaging of LHPs in calcite \cite{ni_observation_2023}; namely, at the bottom of the downward arrow shape, the lenticular shape is connected at $\beta=$~0$^{\circ}$ and thus waves at the surface propagate along $x$ (and all) directions, whereas at higher frequencies, the splitting into two branches indicates that waves at the surface of the crystal will only propagate in some directions. 
Notably, this is consistent with our understanding that larger negative extraordinary permittivity components reduce the directionality associated with ENZ ordinary components (see details of this in Appendix H, in the form of polar plots showing how the LHPs lenticular shape changes with frequency).

When we tilt the anisotropy into the bulk at an angle $\varphi=$~70$^{\circ}$ in Fig.~\ref{fig:leakyazimuthalplots}(b), The intensity of the leaky polariton decreases slightly, and its minimum supporting frequency shifts above ~540~cm\textsuperscript{-1}. This shift is accompanied by an increase in bulk hyperbolic dispersion below ~540~cm\textsuperscript{-1}, illustrating the sensitive dependence of leaky polaritons on the orientation of the crystal's anisotropy.

\section{Discussion}

Our investigation of elliptical surface polaritons, GHPs and LHPs in hyperbolic media using ATR spectroscopy reveals several key insights into their optical behaviour.
We have demonstrated that anisotropy orientation significantly influences the optical properties of hyperbolic polaritons, namely their ATR spectra. 
As such, crystallographic orientation could be of significant importance in designing nanophotonic devices utilising these phenomena, and it offers a new degree of freedom for manipulating light at the nanoscale.

The ATR spectra of GHPs clearly traces the hyperbolic dispersion, outside of the bulk bands, when the anisotropy aligns with the surface, but reflectance reduces in intensity when the anisotropy is tilted with respect to the surface, due to an increase in cross polarisation conversion. This strong direction-dependent behaviour could be used in the directional control of electromagnetic waves. Moreover, GHPs respond to very high in-plane momenta which could be utilised in receiving signals from a wide field of view.
On the other hand, we find that LHPs are limited to a smaller range of $k_x$ values. However, the large asymmetry due to cross polarisation conversion when the anisotropy is oriented away from the surface could be used in novel devices, such as potentially enabling one way communications.

Finally, similar methods to what we explored here could be directly translated to investigating other unusual hyperbolic behaviour such as the newly studied hyperbolic shear polariton in materials with larger asymmetry in their crystal structure \cite{passler_hyperbolic_2022, he_anisotropy_2022, hu_real-space_2023, yves_twist-induced_2024, matson_controlling_2023}. 
In addition, this could also be a way to investigating the impact of twisted-layer structures \cite{yves_twist-induced_2024, wang_strong_2024, orazbay_twistronics-based_2024, sahoo_twisted_2024, alvarez-cuervo_unidirectional_2024, zheng_molding_2022, capote-robayna_twisted_2022, zheng_phonon_2020} on these phenomena, particularly as this could be a way to introduce further control over optical behaviour.

\section{Methods}

To calculate the reflectivity coefficients, the 4x4 Transfer Matrix Method was used \cite{berreman_optics_1972, yeh_electromagnetic_1979, yeh_optical_1990, yeh_optics_1980, schubert_polarization-dependent_1996, passler_layer-resolved_2023, passler_layer-resolved_2020} with full details given in Appendices A and B.
The parameters used to model the optical behaviour of crystal quartz were obtained from the parameters given by Estevam et al.\cite{estevam_da_silva_far-infrared_2012}, built on the work of Gervais and Piraeus \cite{gervais_temperature_1975}.
The experimental data was collected via reflectance measurements using Fourier-transform infrared spectroscopy with a Bruker Vertex 70 Spectrometer. 
The spectra featured a resolution of 4~cm\textsuperscript{-1}, and each spectrum was averaged 15 times.
A KRS-5 polariser was placed in the path of the incident beam to obtain spectra for p-polarised light.
To obtain ATR spectra, a diamond coupling prism was used with a dielectric constant $\epsilon_{p} =$~5.5 and at a fixed incident angle of 45$^{\circ}$.
To introduce an appropriate air gap, 1.5~$\mu$m silica spacers were thinly coated over the ATR stage.
Throughout the azimuthal rotation, the silica spacers were partially moved, so extra care was taken in order to maintain a consistent air gap thickness. 
There is significant margin for error, but results show clear agreement with theory despite this. 
Experimental results align with a theoretical air gap distance of 2~$\mu$m, with variation between 1.5~$\mu$m and 3~$\mu$m (see Appendix C for details of the error associate with the air gap and its effect on reflectance).
The samples used were flat slabs of chemically polished crystal quartz with a 20~mm diameter and 10~mm thickness, provided by Boston Piezo Optics Inc. 
The anisotropy orientation was determined with respect to the crystal's surface using single-crystal X-Ray diffraction, with the crystal mounted on a goniometer for precise positioning at selected orientations.

\section*{Acknowledgement}
MCC acknowledges funding from the Engineering and Physical Sciences Research Council (EPSRC) under grant number EP/S023321/1. We thank N Parry, C A McEleney, M Smith, A Joseph, K Stevens, A MacGruer and S Mekhail for useful insights, discussion and helpful comments on the manuscript.









\section*{Appendix A: The Dielectric Tensor and Anisotropy Orientation}

\renewcommand{\theequation}{A\arabic{equation}}
\renewcommand{\thetable}{A\arabic{table}}
\renewcommand{\thefigure}{A\arabic{figure}}
\setcounter{equation}{0}
\setcounter{table}{0}
\setcounter{figure}{0}
\setcounter{page}{0}

In a uniaxial material like crystal quartz, and when the crystal axis are aligned with the laboratory axes, the dielectric tensor can be written as follows

\begin{equation}
    \epsilon = \begin{bmatrix} 
    \epsilon_{\perp} & 0 & 0 \\ 
    0 & \epsilon_{\perp} & 0 \\ 
    0 & 0 & \epsilon_{\parallel} \\ 
    \end{bmatrix}
\label{eqn:diagonalpermittivitytensor}
\end{equation}
where $\varepsilon_\parallel$ is denotes the
anisotropy axis (in this case along $z$),  
and $\varepsilon_\perp$ denotes the axes perpendicular to the anisotropy, (in our case along $x$ and $y$).

Throughout this paper, we have described the rotation of the anisotropy axis by two angles: $\varphi$ which is the angle between the anisotropy and the crystal's surface and $\beta$ which represents a rotation of the incidence plane with respect to the plane where the anisotropy axis lies (here $x-z$).
Mathematically, these angles correspond to matrix transformations about the $y$ axis and $z$ axis, respectfully.
The rotation of the anisotropy axis by the angle $\varphi$ as described in this work is represented by a matrix transformation around the $y$ axis done by

\begin{equation}
    T_{\varphi} = \begin{bmatrix} 
    \cos{\varphi} & 0 & \sin{\varphi} \\ 
    0 & 1 & 0 \\ 
    -\sin{\varphi} & 0 & \cos{\varphi} \\ 
    \end{bmatrix},
\label{eqn:phirotation}
\end{equation}.

and conversely, the rotation $\beta$ corresponds to a rotation around the $z$ axis given by
\begin{equation}
    T_{\beta} = \begin{bmatrix} 
    \cos{\beta} & \sin{\beta} & 0 \\ 
    -\sin{\beta} & \cos{\beta} & 0 \\ 
    0 & 0 & 1 \\ 
    \end{bmatrix}.
\label{eqn:betarotation}
\end{equation}
The overall rotation matrix is thus given by

\begin{equation}
T = T_{\beta} T_{\varphi},
\end{equation}
and the overall rotated permittivity matrix given by

\begin{equation}
    \epsilon = T \epsilon T^{-1}.
\end{equation}
This results in a 3~$\times$~3 tensor of an arbitrarily oriented anisotropic material, ready to be used in the transfer matrix method:

\begin{equation}
    \epsilon = \begin{bmatrix} 
    \epsilon_{xx} & \epsilon_{xy} & \epsilon_{xz} \\ 
    \epsilon_{yx} & \epsilon_{yy} & \epsilon_{yz} \\ 
    \epsilon_{zx} & \epsilon_{zy} & \epsilon_{zz} \\ 
    \end{bmatrix}.
\label{eqn:permittivitytensorrotation}
\end{equation}

\section*{Appendix B: The Transfer Matrix Method} 

\renewcommand{\theequation}{B\arabic{equation}}
\renewcommand{\thetable}{B\arabic{table}}
\renewcommand{\thefigure}{B\arabic{figure}}
\setcounter{equation}{0}
\setcounter{table}{0}
\setcounter{figure}{0}
\setcounter{page}{0}

The Berreman 4~$\times$~4 Transfer matrix method, 
introduced in 1971 \cite{berreman_optics_1972} 
and later optimised \cite{yeh_electromagnetic_1979, schubert_polarization-dependent_1996, passler_layer-resolved_2020} for more straightforward implementation, calculates eigenvalues for four wave components inside a randomly oriented anisotropic medium, 
with associated eigenvectors providing boundary conditions. 
The method can be easily expanded for multiple layers, 
making it highly useful for ATR simulations. 
Reflection and transmission coefficients can be computed for all polarisations 
$(R_{pp}, R_{ps}, R_{sp}, R_{ss})$, enabling accurate modelling of anisotropic multilayer systems with different anisotropy axis orientations \cite{jones_controlling_2023}. 
Here, we summarise this method which was used throughout the main paper to obtain the theoretical ATR spectra and begin with the definition of a plane wave:

\begin{equation}
    E = E_0 e^{i(kr-\omega t)}
\label{eqn:Eplanewave}
\end{equation}
and
\begin{equation}
    H = H_0 e^{i(kr-\omega t)}.
\label{eqn:Hplanewave}
\end{equation}

Since we are interested in multiple rotations (as outlined in Appendix A), we consider an arbitrary anisotropic material, with fully anisotropic dielectric permittivity and magnetic permeability tensors:

\begin{equation}
    \epsilon_r = \begin{bmatrix} 
    \epsilon_{xx} & \epsilon_{xy} & \epsilon_{xz} \\ 
    \epsilon_{yx} & \epsilon_{yy} & \epsilon_{yz} \\ 
    \epsilon_{zx} & \epsilon_{zy} & \epsilon_{zz} \\ 
    \end{bmatrix}
\label{eqn:permittivitytensor}
\end{equation}
and
\begin{equation}
    \mu_r = \begin{bmatrix} 
    \mu_{xx} & \mu_{xy} & \mu_{xz} \\ 
    \mu_{yx} & \mu_{yy} & \mu_{yz} \\ 
    \mu_{zx} & \mu_{zy} & \mu_{zz} \\ 
    \end{bmatrix}.
\label{eqn:permeabilitytensor}
\end{equation}
We can then employ Maxwell's Equations and the electromagnetic constitutive relations which in CGS units take the following form:

\begin{equation}
    \Delta \times E = -\frac{1}{c} \mu_r\frac{\partial{H}}{\partial{t}}
    \label{eqn:maxwellE}
\end{equation}
and
\begin{equation}
    \Delta \times H = \frac{1}{c} \epsilon_r \frac{\partial{E}}{\partial{t}}.
\label{eqn:maxwellH}
\end{equation}
By solving the partial derivatives [using (\ref{eqn:Eplanewave}) and (\ref{eqn:Hplanewave})] as well as making $\omega/c=k_0$ we get
\begin{equation}
    \Delta \times E = i k_0 \mu_rH
\end{equation}
and
\begin{equation}
    \Delta \times H = - i k_0 \epsilon_r E
\end{equation}
We can also say that $\frac{\partial}{\partial{x}} \equiv ik_x$ and $\frac{\partial}{\partial{y}} \equiv ik_y$ and since there is no contribution from the y-component of the wavevector, $\frac{\partial}{\partial{y}} \equiv 0$.
We will not extend this to the $z$-component, as the sample will not be homogenous in $z$. Therefore we will keep $\frac{\partial}{\partial{z}}$ the way it is.
From this point on, we will normalise our spatial components by a factor of $k_0$ whereby:

\begin{equation}
    K_x \equiv \frac{k_x}{k_0}
\label{eqn:Kx}
\end{equation} which is known as the reduced wavevector, and
\begin{equation}
    z' \equiv k_0z
\end{equation}
so that
\begin{equation}
    \frac{\partial}{\partial{z'}} \equiv \frac{1}{k_0} \frac{\partial}{\partial{z}}
\end{equation}
This leaves us with:
\begin{equation}
    \Delta \times E = i \mu_rH
\label{eqn:DeltaE}
\end{equation}
and
\begin{equation}
    \Delta \times H = - i\epsilon_r E.
\label{eqn:DeltaH}
\end{equation}

Following from Eq.~(\ref{eqn:DeltaE}) we get
\begin{equation}
    \begin{vmatrix} 
    i & j & k \\ 
    iK_x & 0 & \frac{\partial}{\partial{z'}} \\ 
    E_x & E_y & E_z 
    \end{vmatrix} = i\begin{bmatrix} 
    \mu_{xx} & \mu_{xy} & \mu_{xz} \\ 
    \mu_{yx} & \mu_{yy} & \mu_{yz} \\ 
    \mu_{zx} & \mu_{zy} & \mu_{zz} \\ 
    \end{bmatrix}
    \begin{bmatrix}
    H_x \\ H_y \\ H_z
    \end{bmatrix}
\end{equation}
which gives us three linear equations:
\begin{equation}
    \frac{\partial}{\partial z'} E_y = -i (\mu_{xx} H_x + \mu_{xy} H_y + \mu_{xz} H_z),
\label{eqn:DeltaE1Part1}
\end{equation}
\begin{equation}
    \frac{\partial}{\partial z'} E_x - i K_x E_z = i(\mu_{yx} H_x + \mu_{yy} H_y + \mu_{yz} H_z),
\label{eqn:DeltaE1Part2}
\end{equation}
and
\begin{equation}
    iK_xE_y = i(\mu_{zx}H_x + \mu_{zy}H_y + \mu_{zz}H_z).
\label{eqn:DeltaE1Part3}
\end{equation}
These equations can then be rearranged as
\begin{equation}
\frac{1}{i}\frac{\partial}{\partial z'} E_y = -(\mu_{xx} H_x + \mu_{xy} H_y + \mu_{xz} H_z),
\label{eqn:DeltaE1}
\end{equation}
\begin{equation}
\frac{1}{i}\frac{\partial}{\partial z'} E_x = K_x E_z + \mu_{yx} H_x + \mu_{yy} H_y + \mu_{yz} H_z,
\label{eqn:DeltaE2}
\end{equation}
and
\begin{equation}
H_z = \frac{1}{\mu_{zz}}(K_x E_y - \mu_{zx} H_x - \mu_{zy} H_y).
\label{eqn:DeltaE3}
\end{equation}

Repeating this process for Eq.~(\ref{eqn:DeltaH})
yields
\begin{equation}
\frac{1}{i}\frac{\partial}{\partial z'} H_y = \epsilon_{xx} H_x + \epsilon_{xy} H_y + \epsilon_{xz} E_z,
\label{eqn:DeltaH1}
\end{equation}
\begin{equation}
\frac{1}{i}\frac{\partial}{\partial z'} H_x = K_x H_z - \epsilon_{yx} E_x - \epsilon_{yy} E_y - \epsilon_{yz} E_z,
\label{eqn:DeltaH2}
\end{equation}
and
\begin{equation}
E_z = -\frac{1}{\epsilon_{zz}}(K_x H_y + \epsilon_{zx} E_x + \epsilon_{zy} E_y).
\label{eqn:DeltaH3}
\end{equation}
With this, we now have expressions for $E_z$ and $H_z$ in terms of $E_x$, $E_y$, $H_x$ and $H_y$, along
with four equations for the partial derivatives of $E_x$, $E_y$, $H_x$ and $H_y$ with respect to $z'$ 
in terms of $E_z$ and $H_z$.

Now we can subsitute $E_z$ and $H_z$ in to Eqs.~(\ref{eqn:DeltaE1}), (\ref{eqn:DeltaE2}), (\ref{eqn:DeltaH1}), (\ref{eqn:DeltaH2}).
We begin with Eq.~(\ref{eqn:DeltaE2})
\begin{equation}
    \begin{split}
        \frac{1}{i}\frac{\partial}{\partial z'} E_x = 
        K_x \left[-\frac{1}{\epsilon_{zz}}(K_x H_y + \epsilon_{zx} E_x + \epsilon_{zy} E_y)\right] + \\
        \mu_{yx} H_x + \mu_{yy} H_y + \mu_{yz} \left[\frac{1}{\mu_{zz}}(K_x E_y - \mu_{zx} H_x - \mu_{zy} H_y)\right]
    \end{split}
\end{equation}
which after substantial rearranging, yields
\begin{equation}
    \begin{split}
        \frac{1}{i}\frac{\partial}{\partial z'} E_x 
        = 
        & E_x\left[-K_x\left(\frac{\epsilon_{zx}}{\epsilon_{zz}}\right)\right] \\
        & + E_y\left[K_x\left(\frac{\mu_{yz}}{\mu_{zz}} - \frac{\epsilon_{zy}}{\epsilon_{zz}}\right)\right] \\
        & + H_x\left(\mu_{yx} - \frac{\mu_{yz}\mu_{zx}}{\mu_{zz}}\right) \\
        & + H_y\left(\mu_{yy} - \frac{\mu_{yz}\mu_{zy}}{\mu_{zz}} - \frac{K_x \epsilon_{zy}}{\epsilon_{zz}}\right).
    \end{split}
\label{eqn:ExPartial}
\end{equation}
Eq.~(\ref{eqn:DeltaE1}) becomes

\begin{equation}
    \begin{split}
        \frac{1}{i}\frac{\partial}{\partial z'} E_y 
        &= -\mu_{xx} H_x - \mu_{xy} H_y \\
        &\phantom{=} -\frac{\mu_{xz}}{\mu_{zz}}(K_x E_y - \mu_{zx} H_x - \mu_{zy} H_y)
    \end{split}
\end{equation}
which after rearranging yields
\begin{equation}
    \begin{split}
    \frac{1}{i}\frac{\partial}{\partial z'} E_y
    = 
    &E_y\left(-\frac{K_x \mu_{xz}}{\mu_{zz}}\right)\\
    +
    &H_x\left(\frac{\mu_{xz}\mu_{zx}}{\mu_{zz}} - \mu_{xx}\right)\\
    +
    &H_y\left(\frac{\mu_{xz}\mu_{zy}}{\mu_{zz}} - \mu_{xy}\right).
    \end{split}
\label{eqn:EyPartial}
\end{equation}
Eq.~(\ref{eqn:DeltaH2}) now is
\begin{equation}
    \begin{split}
        \frac{1}{i}\frac{\partial}{\partial z'} H_x
        = & K_x \left[\frac{1}{\mu_{zz}}(K_x E_y - \mu_{zx} H_x - \mu_{zy} H_y)\right] \\
        & - \epsilon_{yx} E_x - \epsilon_{yy} E_y \\
        & - \epsilon_{yz} \left[-\frac{1}{\epsilon_{zz}}(K_x H_y + \epsilon_{zx} E_x + \epsilon_{zy} E_y)\right]
    \end{split}
\end{equation}
which and after rearranged take the form
\begin{equation}
    \begin{split}
        \frac{1}{i}\frac{\partial}{\partial z'} H_x 
        = & E_x\left(\frac{\epsilon_{yz}\epsilon_{zx}}{\epsilon_{zz}} - \epsilon_{yx}\right) \\
        & + E_y\left(\frac{K_x^2}{\mu_{zz}} + \frac{\epsilon_{yz}\epsilon_{zy}}{\epsilon_{zz}} - \epsilon_{yy}\right) \\
        & + H_x\left(-\frac{K_x \mu_{zx}}{\mu_{zz}}\right) \\
        & + H_y\left[K_x \left(\frac{\epsilon_{yz}}{\epsilon_{zz}} - \frac{\mu_{zy}}{\mu_{zz}}\right)\right].
    \end{split}
\label{eqn:HxPartial}
\end{equation}
Finally, Eq.~(\ref{eqn:DeltaH1}) can be written as
\begin{equation}
    \begin{split}
        \frac{1}{i}\frac{\partial}{\partial z'} H_y
        &= \epsilon_{xx} H_x + \epsilon_{xy} H_y \\
        &\phantom{=} + \epsilon_{xz} \left[-\frac{1}{\epsilon_{zz}}(K_x H_y + \epsilon_{zx} E_x + \epsilon_{zy} E_y)\right],
    \end{split}
\end{equation}
and after rearranging it becomes
\begin{equation}
    \begin{split}
        \frac{1}{i}\frac{\partial}{\partial z'} H_y
        &= E_x\left(\epsilon_{xx} - \frac{\epsilon_{xz} \epsilon_{zx}}{\epsilon_{zz}}\right) \\
        &\phantom{=} + E_y\left(\epsilon_{xy} - \frac{\epsilon_{xz} \epsilon_{zy}}{\epsilon_{zz}}\right) \\
        &\phantom{=} + H_y\left(-\frac{K_x \epsilon_{xz}}{\epsilon_{zz}}\right).
    \end{split}
\label{eqn:HyPartial}
\end{equation}

With Eqs.~(\ref{eqn:ExPartial}), (\ref{eqn:EyPartial}), (\ref{eqn:HxPartial}), and (\ref{eqn:HyPartial}), we have four linearly independent equations in terms of the partial $z'$ derivative for each component of an incident wave of arbitrary polarisation.
These four equations can be written in the form:
\begin{equation}
    \frac{\partial}{\partial z'}
    \begin{bmatrix} 
    E_x\\ E_y\\  H_x\\ H_y
    \end{bmatrix}
    =
    i \Delta
    \begin{bmatrix} 
    E_x\\ E_y\\  H_x\\ H_y
    \end{bmatrix}
\label{eqn:initialMatrixConstruction}
\end{equation}
where $\Delta$ is a 4~$\times$~4 matrix:
{
\begin{equation}
    \Delta \equiv
    \begin{bmatrix} 
    -K_x(\frac{\epsilon_{zx}}{\epsilon_{zz}}) 
    & 
    K_x(\frac{\mu_{yz}}{\mu_{zz}} - \frac{\epsilon_{zy}}{\epsilon_{zz}}) 
    & 
    \mu_{yx} - \frac{\mu_{yz}\mu_{zx}}{\mu_{zz}} 
    &
    \mu_{yy} - \frac{\mu_{yz}\mu_{zy}}{\mu_{zz}} - \frac{K_x \epsilon_{zy}}{\epsilon_{zz}}
    \\
    0 
    &
    -\frac{K_x \mu_{xz}}{\mu_{zz}}
    &
    \frac{\mu_{xz}\mu_{zx}}{\mu_{zz}} - \mu_{xx}
    &
    \frac{\mu_{xz}\mu_{zy}}{\mu_{zz}} - \mu_{xy}
    \\
    \frac{\epsilon_{yz}\epsilon_{zx}}{\epsilon_{zz}} - \epsilon_{yx}
    &
    \frac{K_x^2}{\mu_{zz}} + \frac{\epsilon_{yz}\epsilon_{zy}}{\epsilon_{zz}} - \epsilon_{yy}
    &
    -\frac{K_x \mu_{zx}}{\mu_{zz}}
    &
    K_x (\frac{\epsilon_{yz}}{\epsilon_{zz}} - \frac{\mu_{zy}}{\mu_{zz}})
    \\
    \epsilon_{xx} - \frac{\epsilon_{xz} \epsilon_{zx}}{\epsilon_{zz}}
    & 
    \epsilon_{xy} - \frac{\epsilon_{xz} \epsilon_{zy}}{\epsilon_{zz}}
    & 
    0
    &
    -\frac{K_x \epsilon_{xz}}{\epsilon_{zz}} 
    \end{bmatrix}
\label{eqn:BerremanDeltaMatrix}
\end{equation}
}

For a nonmagnetic medium (or material without an anisotropic magnetic permeability tensor), this will be: 

The overall partial transfer matrix $T$ for a given material of arbitrary thickness $d$ is then given by

\begin{equation}
    T = \exp{\left(i k_0 \Delta d\right)}.
\label{eqn:PartialMatrix}
\end{equation}
As our wavenumber $\omega$ is measured in cm\textsuperscript{-1}, the thickness $d$ is given in cm for the purpose of our analysis.
When taking the exponential of the 4~$\times$~4 matrix, it would be too computationally intensive to calculate the exponential of every element. 
Instead, we can find the eigenvalues and corresponding column eigenvectors to apply the exponential function:
\begin{equation}
    \exp{\left(\Delta\right)} = V e^{\lambda} V^{-1}
\label{eqn:ExponentialMatrix}
\end{equation}
where $V$ is the matrix of eigenvectors and $\lambda$ is the diagonal matrix of eigenvalues.
The eigenvalues will be four values, with the eigenvectors a 4~$\times$~4 matrix, where each $n$th column is the column eigenvector for the $n$th eigenvalue.

Therefore, the transfer matrix for a given material is given by:
\begin{equation}
    T = V \exp{\left(i \lambda k_0 d\right)} V^{-1}
\label{eqn:FinalTransferMatrix}
\end{equation}
where the eigenvectors $V$ correspond to the boundary conditions at the entry interface of the medium. 
The eigenvalues $\lambda$ tell us the propagation properties of each mode inside the medium. 
The inverse of the eigenvectors $V^{-1}$ correspond to the boundary conditions at the exit interface of the medium.

By repeating this method for all the films in a system, 
an overall transfer matrix can be found by multiplying each 
partial transfer matrix in the corresponding order of each medium.
A full transfer matrix can be made by projecting the fields into the ambient and substrate mediums, given by:
\begin{equation}
    \bar{T} = L_i^{-1} T L_t
\end{equation}
where
\begin{equation}
    {\renewcommand\arraystretch{1.2}
    \scalebox{1.2}{$L_i^{-1} =
    \begin{bmatrix}
    0 & 1 & -\frac{1}{n\cos\theta_i} & 0 \\
    0 & 1 & \frac{1}{n\cos\theta_i} & 0 \\
    \frac{1}{\cos\theta_i} & 0 & 0 & \frac{1}{n} \\
    -\frac{1}{\cos\theta_i} & 0 & 0 & \frac{1}{n}
    \end{bmatrix}$}}
\end{equation}
and
\begin{equation}
    {\renewcommand\arraystretch{1.2}
    \scalebox{1.2}{$L_t =
    \begin{bmatrix}
        0 & 0 & A & -A \\
        1 & 1 & 0 & 0 \\
        -A & n_sA & 0 & 0\\
        0 & 0 & n_s & n_s
    \end{bmatrix}$}}
\end{equation}
where $A = \cos\theta_t$ 
\cite{schubert_polarization-dependent_1996}.

When we wish to model our final layer as semi-infinite, 
we do not extend our transfer matrix in the same way as we do with an isotropic substrate, as there is no singular refractive index value that we can use.
To refresh, the fields in our system can be expressed as follows
\begin{equation}
    \begin{bmatrix} 
    A_s\\ B_s\\  A_p\\ B_p
    \end{bmatrix}
    = T
    \begin{bmatrix} 
    C_s\\ D_s\\  C_p\\ D_p
    \end{bmatrix}
\end{equation}
where $s$ and $p$ denote polarisations, $A$ denotes positive traveling waves in the initial medium, $B$ denotes backward waves in the initial medium, with $C$ and $D$ denoting the respective counterparts in the exit medium.
When calculating the 4~$\times$~4 Transfer Matrix, the column eigenvectors are the boundary conditions of the layer for each mode of propagation, represented by each eigenvalue. 
For a semi-infinite layer, we assume that there are no backward-traveling waves, which means that the $D_s = D_p = 0$.
To achieve this, we need to examine the four eigenvalues of the matrix. Two eigenvalues correspond to forward waves, and the other two correspond to backward waves. 
We must sort these eigenvalues by their imaginary components, so that evanescent waves are decaying exponentially from the interface, to satisfy the conservation of energy. 
We also extract the two column eigenvectors linked with the positive-traveling eigenvalues. 
We then arrange these eigenvectors in the first and third columns of the matrix so that they can be properly multiplied by $C_s$ and $C_p$. 
The remaining two eigenvectors related to backward-traveling waves are discarded.
Now, the partial transfer matrix of an anisotropic semi-infinite layer can be expressed as a 4~$\times$~4 matrix, where the first and third columns contain the `positive' eigenvectors, while the second and fourth columns are set equal to zero.

Now that the overall transfer matrix is found for the full system needed for our ATR investigation, which can be comprised
of an arbitrary number of anisotropic layers or arbitrary anisotropy axis orientation, the reflection and transmission coefficients
of the whole system can easily be found using
\begin{equation}
    \begin{bmatrix} 
        A_s\\ B_s\\  A_p\\ B_p
    \end{bmatrix}
    =
    \begin{bmatrix} 
    T_{11} & T_{12} & T_{13} & T_{14}
    \\
    T_{21} & T_{22} & T_{23} & T_{24}
    \\
    T_{31} & T_{32} & T_{33} & T_{34}
    \\
    T_{41} & T_{42} & T_{43} & T_{44}
    \end{bmatrix}
    \begin{bmatrix} 
    C_s\\ 0\\  C_p\\ 0
    \end{bmatrix}.
\end{equation}
In this work, we are most interested in the reflection coefficients, and for all polarisations from this Transfer Matrix these are
$r_{pp}, r_{ps}, r_{sp}, r_{ss}$.
Namely, $r_{pp}$  represents the ratio of p-polarised light exiting 
the medium to p-polarised light entering the medium, i.e. when $A_s = 0$.
As a function of the transfer matrix components, $r_{pp}$ is given by
\begin{equation}
r_{pp} = \frac{B_p}{A_p}_{A_s = 0},
\end{equation}
where
\begin{equation}
A_s = M_{11}C_s + M_{13} C_p = 0,
\end{equation}
\begin{equation}
C_s = -\frac{M_{13}}{M_{11}} C_p,
\end{equation}
\begin{equation}
B_p = M_{41} C_s + M_{43} C_p,
\end{equation}
and
\begin{equation}
A_p = M_{31}C_s + M_{33}C_p.
\end{equation}
Substituting $C_s$ in to $B_p$ and $A_p$ we obtain
\begin{equation}
r_{pp} = \frac{B_p}{A_p}_{A_s = 0} =\frac{M_{11}M_{43} - M_{41}M_{13}}{M_{11}M_{33} - M_{13}M_{31}}
\label{eqn:r_pp}
\end{equation}

By similar analysis, we can find the other reflection coefficients.
For instance, $r_{ss} $ represents the ratio of s-polarised light exiting the medium to s-polarised light entering the medium, i.e. when $A_p = 0$ and it is given by
\begin{equation}
r_{ss} = \frac{B_s}{A_s}_{A_p = 0} =\frac{M_{21}M_{33} - M_{23}M_{31}}{M_{11}M_{33} - M_{13}M_{31}}.
\label{eqn:r_ss}
\end{equation}
Similarly, $r_{sp} $ represents the ratio of p-polarised light exiting the medium to s-polarised light entering the medium, i.e. when $A_p = 0$ and it is given by
\begin{equation}
r_{sp} = \frac{B_p}{A_s}_{A_p = 0} =\frac{M_{41}M_{33} - M_{43}M_{31}}{M_{11}M_{33} - M_{13}M_{31}}.
\label{eqn:r_sp}
\end{equation}
Finally, $r_{ps} $ represents the ratio of s-polarised light exiting
 the medium to p-polarised light entering the medium, i.e. when $A_s = 0$ and it can be written as
\begin{equation}
r_{ps} = \frac{B_s}{A_p}_{A_s = 0} =\frac{M_{11}M_{23} - M_{21}M_{13}}{M_{11}M_{33} - M_{13}M_{31}}.
\label{eqn:r_ps}
\end{equation}

With these coefficients, the reflectance for each polarisation can be calculated as
\begin{equation}
        R_{pp} = r_{pp} r_{pp}^*,
\label{eqn:Rpp}
\end{equation}
\begin{equation}
    R_{ps} = r_{ps} r_{ps}^*,
\label{eqn:Rps}
\end{equation}
\begin{equation}
    R_{sp} = r_{sp} r_{sp}^*,
\label{eqn:Rsp}
\end{equation}
and
\begin{equation}
    R_{ss} = r_{ss} r_{ss}^*.
\label{eqn:Rss}
\end{equation}

We can also define quantities
\begin{equation}
    R_p = R_{pp} + R_{ps}
\label{R_p}
\end{equation}
and
\begin{equation}
    R_s = R_{sp} + R_{ss}
\label{R_s}
\end{equation}
which represent the total reflectance of p-polarised and s-polarised incident radiation, 
respectively.
Note that throughout the main body of our work, our reflectivity is given by the quantity $R_p$ as we only considered the case of p-polarised incident radiation.

\section*{Appendix C: Critical Coupling by varying Air Gap}

\renewcommand{\theequation}{C\arabic{equation}}
\renewcommand{\thetable}{C\arabic{table}}
\renewcommand{\thefigure}{C\arabic{figure}}
\setcounter{equation}{0}
\setcounter{table}{0}
\setcounter{figure}{0}
\setcounter{page}{0}

\begin{figure*}[!ht]
\centering
\includegraphics[width=\linewidth]{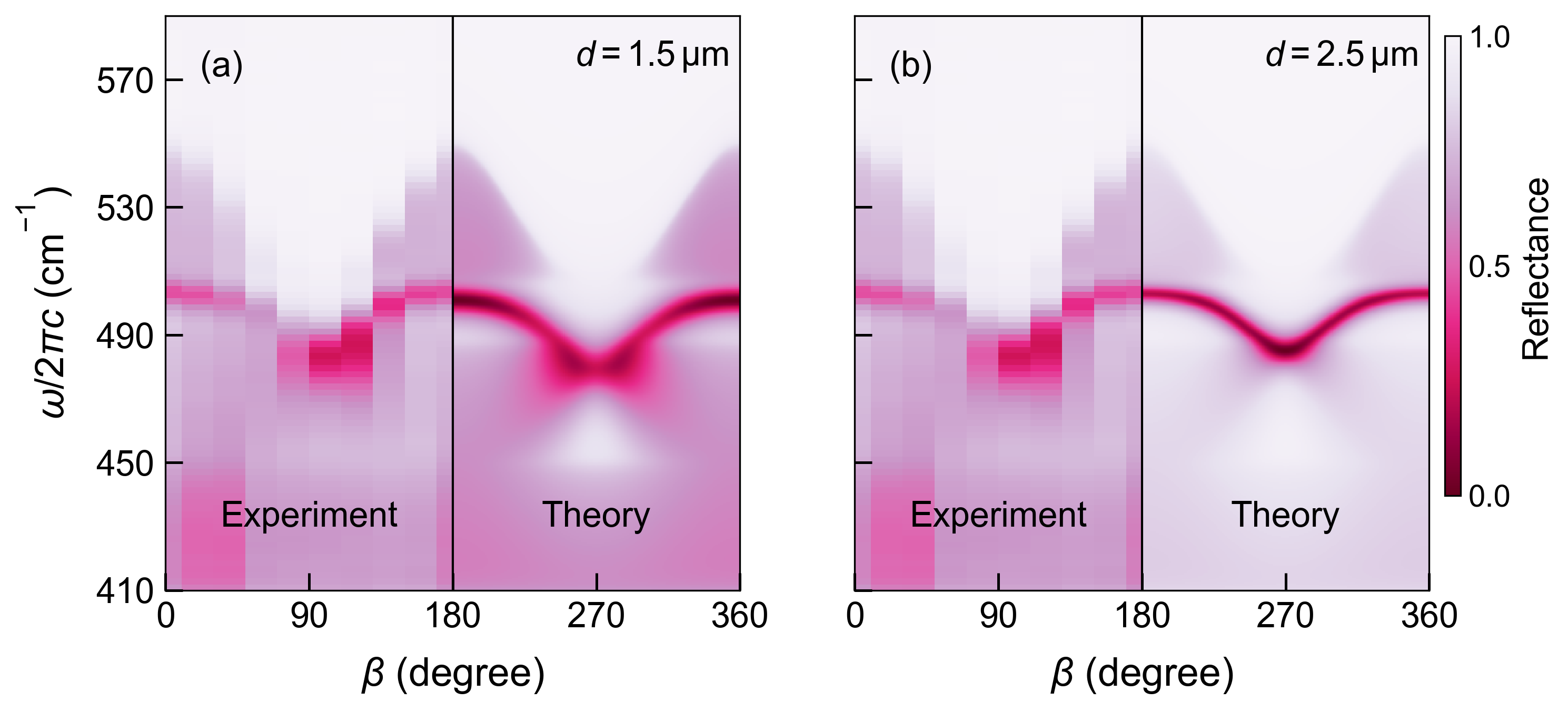}
\caption{The dependence of ATR in crystal quartz on the azimuthal angle $\beta$ when $\varphi=$~90$^{\circ}$ and $\frac{k_x}{k_0} = 1.66$.
We show our experimental work alongside two different theoretical air gap sizes, where in a) $d =$~1.5~$\mu$m. and in b) $d =$~2.5~$\mu$m.
}
\label{fig:experimentalairgapcomparison}
\end{figure*}

For our experimental data (Figs. 1 and 3 of the main article), we selected a theoretical air gap of $d =$~2~$\mu$m to compare with experimental ATR results.
In Fig.~\ref{fig:experimentalairgapcomparison}, we show two other air gap sizes [in Fig.~\ref{fig:experimentalairgapcomparison}(a) $d =$~1.5~$\mu$m and in Fig.~\ref{fig:experimentalairgapcomparison}(b) $d =$~2.5~$\mu$m]
We can see how in both the elliptical surface wave is slightly not matching with the experimental result. 
In (a) the peak frequency at $\beta=$~180$^{\circ}$ is slightly lower in the theoretical results compared with the experimental while in (b) the coupling of the surface polariton is much stronger and overall freqeuncies match well the experiment, however it is coupled slightly more than the experimental value as evidenced by higher reflectance at bulk freqeuncies.
As these are qualitative observations, it was deemed that $d =$~2~$\mu$m was most appropriate for inclusion within the main body of this paper as it is a good match for both freqeuncy of the surface mode and overall intensity matching.

\section*{Appendix D: Cross Polarisation Conversion of the Ghost Hyperbolic Polariton}

\renewcommand{\theequation}{D\arabic{equation}}
\renewcommand{\thetable}{D\arabic{table}}
\renewcommand{\thefigure}{D\arabic{figure}}
\setcounter{equation}{0}
\setcounter{table}{0}
\setcounter{figure}{0}
\setcounter{page}{0}

\begin{figure*}[!ht]
\centering
\includegraphics[width=\linewidth]{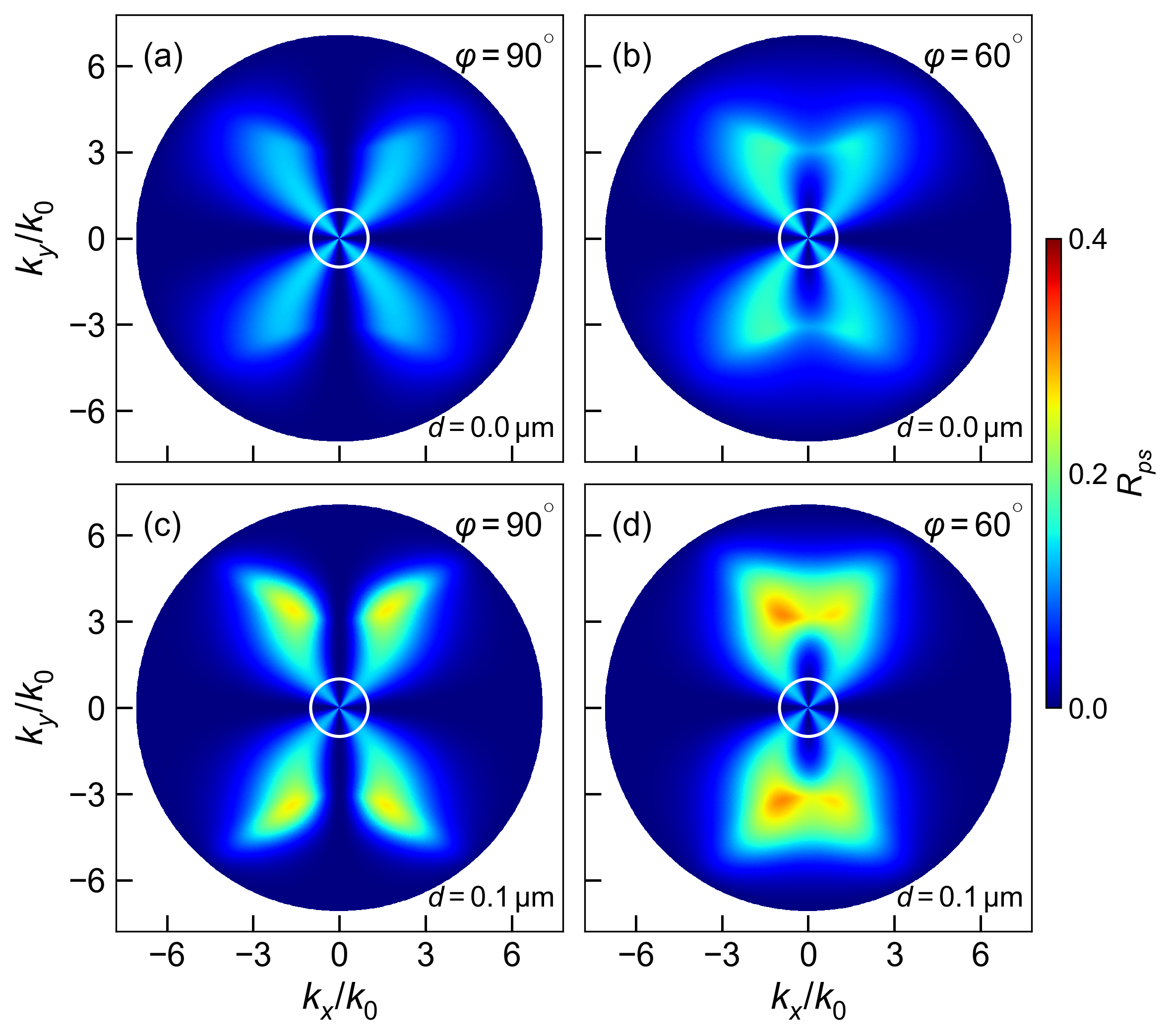}
\caption{
The cross-polarisation conversion ($R_{ps}$) induced by the GHP in Quartz at $\omega/2\pi c=$~460~cm\textsuperscript{-1}. This quantity is the amount of s-polarised light reflected off the surface from incident p-polarised radiation.
The radius corresponding to $k_x$ where $\epsilon_p =$~50, and the azimuthal angle corresponding to the angle $\beta$.
The white circle denotes where $k_x/k_0 = 1$.
In a) there is no air gap ($d =$~0~$\mu$m) and $\varphi=$~90$^{\circ}$.
In b) the anisotropy orientation is unchanged ($\varphi=$~90$^{\circ}$) and an air gap is introduced to study the GHP, with $d =$~0.1~$\mu$m.
In c), the air gap is removed ($d =$~0~$\mu$m) and the anisotropy is rotated to $\varphi=$~60$^{\circ}$ to alter the hyperbolic dispersion.
In d), the anisotropy orientation is unchanged ($\varphi=$~60$^{\circ}$) and an air gap is introduced again to study the GHP, with $d =$~0.1~$\mu$m.
}
\label{fig:ghostcrosspolarisation}
\end{figure*}

Asymmetric cross polarisation conversion has previously been observed in bulk Type I hyperbolic dispersion \cite{wu_asymmetric_2022}. In Fig.~\ref{fig:ghostcrosspolarisation}, we show the value $R_{ps}$, the reflected radiation that possesses s-polarisation from incident p-polarised radiation, associated with Type II hyperbolic dispersion and the GHP.
In Fig.~\ref{fig:ghostcrosspolarisation}(a),where there is no air gap ($d =$~0~$\mu$m) and $\varphi=$~90$^{\circ}$, we can see minimal cross polarisation conversion, at a maximum in an "X" shape at ~45$^{\circ}$ intervals of the angle $\beta$.
When we tilt the anisotropy to $\varphi=$~60$^{\circ}$ in Fig.~\ref{fig:ghostcrosspolarisation}(b), asymmetry is introduced in the cross polarisation conversion, with negative $k_x$ values possessing $R_{ps}$ values around ~0.15, in contrast with positive $k_x$ values possessing $R_{ps}$ values around ~0.1.

We now introduce the air gap of $d =$~0.1~$\mu$m to study cross polarisation conversion induced by the GHP.
In Fig.~\ref{fig:ghostcrosspolarisation}(c), where $\varphi=$~90$^{\circ}$, the maximum $R_{ps}$ has increased to around ~0.27, in a more pronounced symmetric "X" shape.
Tilting the anisotropy to $\varphi=$~60$^{\circ}$ in Fig.~\ref{fig:ghostcrosspolarisation}(d), $R_{ps}$ becomes much more noticeably asymmetric, reaching a maximum of around ~0.36 for negative $k_x$ values, and a maximum of around ~0.26 for positive $k_x$ values. Moreover, the pronounced "X" shape has compressed inwards, joining slightly at $k_x/k_0 = $ ~0.

\section*{Appendix E: Air Gap Dependence of the Ghost Hyperbolic Polariton}

\renewcommand{\theequation}{E\arabic{equation}}
\renewcommand{\thetable}{E\arabic{table}}
\renewcommand{\thefigure}{E\arabic{figure}}
\setcounter{equation}{0}
\setcounter{table}{0}
\setcounter{figure}{0}
\setcounter{page}{0}

\begin{figure}[!ht]
\centering
\includegraphics[width = \linewidth, keepaspectratio]{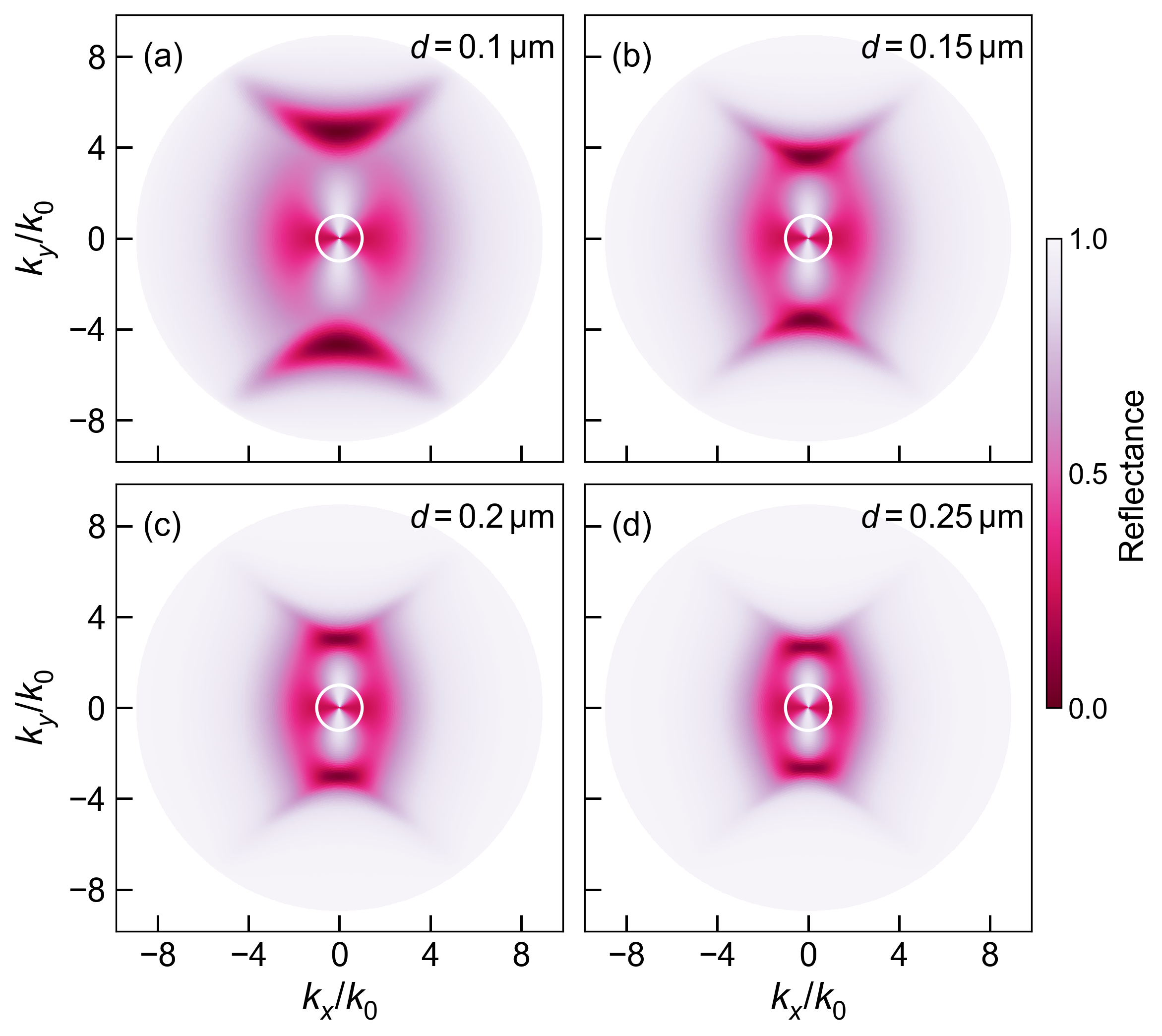}
\caption{
The dependency of the ATR response of Quartz on the air gap thickness $d$ at constant frequency $\omega/2\pi c=$~465~cm\textsuperscript{-1} and the anisotropy axis aligned with the 
surface ($\varphi=$~90$^{\circ}$), with the radius corresponding to $k_x$ 
where $\epsilon_p =$~80, and the azimuthal angle corresponding to the angle $\beta$.
The white circle denotes where $k_x/k_0=$~1~.
In a) $d =$~0.1~$\mu$m.
In b) $d =$~0.15~$\mu$m.
In c), $d =$~0.2~$\mu$m.
In d), $d =$~0,25~$\mu$m.
}

\label{fig:GhostAirGap}
\end{figure}

As outlined previously in Appendix C, the chosen air gap size affects the coupling with the surface polariton due to the decay length of evanescent waves at the prism/air interface.

In Fig.~\ref{fig:GhostAirGap}, we show how changing the air gap affects the reflectance associated with the GHP in quartz at $\omega/2\pi c=$~465~cm\textsuperscript{-1}, when the anisotropy is aligned with the surface ($\varphi=$~90$^{\circ}$).
In Fig.~\ref{fig:GhostAirGap}(a) $d =$~0.1~$\mu$m and the GHP is clearly tracing the bulk hyperbolic dispersion. At $k_x/k_0$ = ~0, the GHP's drop in reflectance is quite thick, supported for $k_y/k_0$ values between ~4 and ~6.
We increase the air gap thickness to $d =$~0.15~$\mu$m in Fig.~\ref{fig:GhostAirGap}(b), where the GHPs signature hyperbolic shape becomes thinner and moves in closer to the bulk dispersion, now supported for $k_y/k_0$ values between ~3.5 and ~4.5 when $k_x/k_0$ = ~0. The four directional lobes are slightly more prominent here, but supported at smaller $k$ values than Fig.~\ref{fig:GhostAirGap}(a).
Increasing the air gap thickness further to $d =$~0.2~$\mu$m in Fig.~\ref{fig:GhostAirGap}(c) begins to diminish the reflective presence of the GHP. At $k_x/k_0$ = ~0, the GHP is supported for $k_y/k_0$ values between 3 and 4, overlapping with the bulk hyperbola, where bulk propagation was forbidden with no air gap as shown in Fig.~\ref{fig:ghostpolarplots}(a) in the main paper.
Increasing the air gap to $d =$~0.25~$\mu$m in Fig.~\ref{fig:GhostAirGap}(d) removes most traces of the GHP's reflective footprint, except for the drop in reflectivity at $k_x/k_0=$~0.

\section*{Appendix F: Frequency Dependence of the Ghost Hyperbolic Polariton}

\renewcommand{\theequation}{F\arabic{equation}}
\renewcommand{\thetable}{F\arabic{table}}
\renewcommand{\thefigure}{F\arabic{figure}}
\setcounter{equation}{0}
\setcounter{table}{0}
\setcounter{figure}{0}
\setcounter{page}{0}

\subsection{Quartz}

\begin{figure}[!ht]
\centering
\includegraphics[width = \linewidth, keepaspectratio]{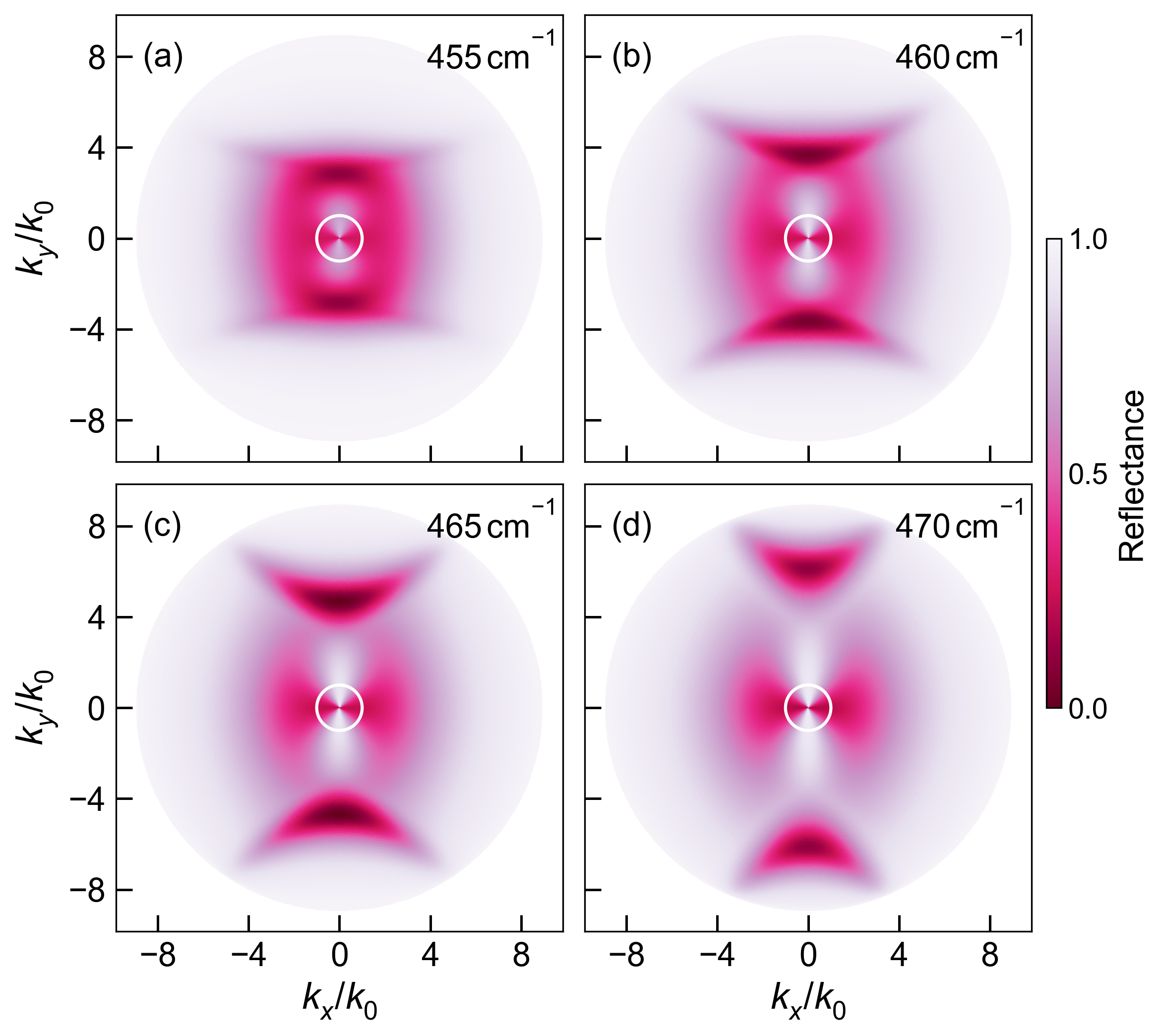}
\caption{
The frequency dependent ATR response of Quartz at with a constant air gap $d =$~0.1~$\mu$m and the anisotropy axis aligned with the 
surface ($\varphi=$~90$^{\circ}$), with the radius corresponding to $k_x$ 
where $\epsilon_p =$~80, and the azimuthal angle corresponding to the angle $\beta$.
The white circle denotes where $k_x/k_0=$~1~.
In a) $\omega/2\pi c=$~455~cm\textsuperscript{-1}.
In b) $\omega/2\pi c=$~460~cm\textsuperscript{-1}.
In c), $\omega/2\pi c=$~465~cm\textsuperscript{-1}.
In d), $\omega/2\pi c=$~470~cm\textsuperscript{-1}.
}
\label{fig:QuartzGhostFrequency}
\end{figure}

We have shown how the GHP can be supported in the Type II hyperbolic region in Quartz. However, compared to Calcite where the positive $\varepsilon_{\parallel}$ is constant through the Type II region, in Quartz both $\varepsilon_{\parallel}$ and $\varepsilon_{\perp}$ vary greatly throughout the hyperbolic band. 
With a constant air gap $d =$~0.1~$\mu$m and the anisotropy axis aligned with the surface ($\varphi=$~90$^{\circ}$), in Fig.~\ref{fig:QuartzGhostFrequency}, we show how the GHP is supported at different frequencies throughout the Type II hyperbolic region in Quartz, from ~455~cm\textsuperscript{-1} to ~470~cm\textsuperscript{-1}. As the frequency increases within this range, the magnitude of the positive $\varepsilon_{\parallel}$ increases from ~8 to ~13.5, while the negative $\varepsilon_{\perp}$ shrinks from ~-27 to ~-6.
In Fig.~\ref{fig:QuartzGhostFrequency}(a) at ~455~cm\textsuperscript{-1}, the hyperbolic dispersion is quite flat, so the GHP is not very clear, but can be slightly observed emanating outwards as faint horizontal lines at $k_y/k_0$ = ~4.
In Fig.~\ref{fig:QuartzGhostFrequency}(b) at ~460~cm\textsuperscript{-1} the GHP is much more prevalent, as observed in the main paper.
Increasing the frequency further to ~465~cm\textsuperscript{-1} in Fig.~\ref{fig:QuartzGhostFrequency}(c), the hyperbolic dispersion widens, causing the GHP to be supported at narrower $k_x$ values, but larger $k_y$ values.
This trend continues with increasing the frequency to ~470~cm\textsuperscript{-1} in Fig.~\ref{fig:QuartzGhostFrequency}(d), with a much narrow reflective footprint, with a smaller dip in reflectance compared to (b) and (c).
These results suggest how the GHP can isolate specific frequencies depending on crystal orientation.

\subsection{Calcite}

\begin{figure}[!ht]
\centering
\includegraphics[width = \linewidth, keepaspectratio]{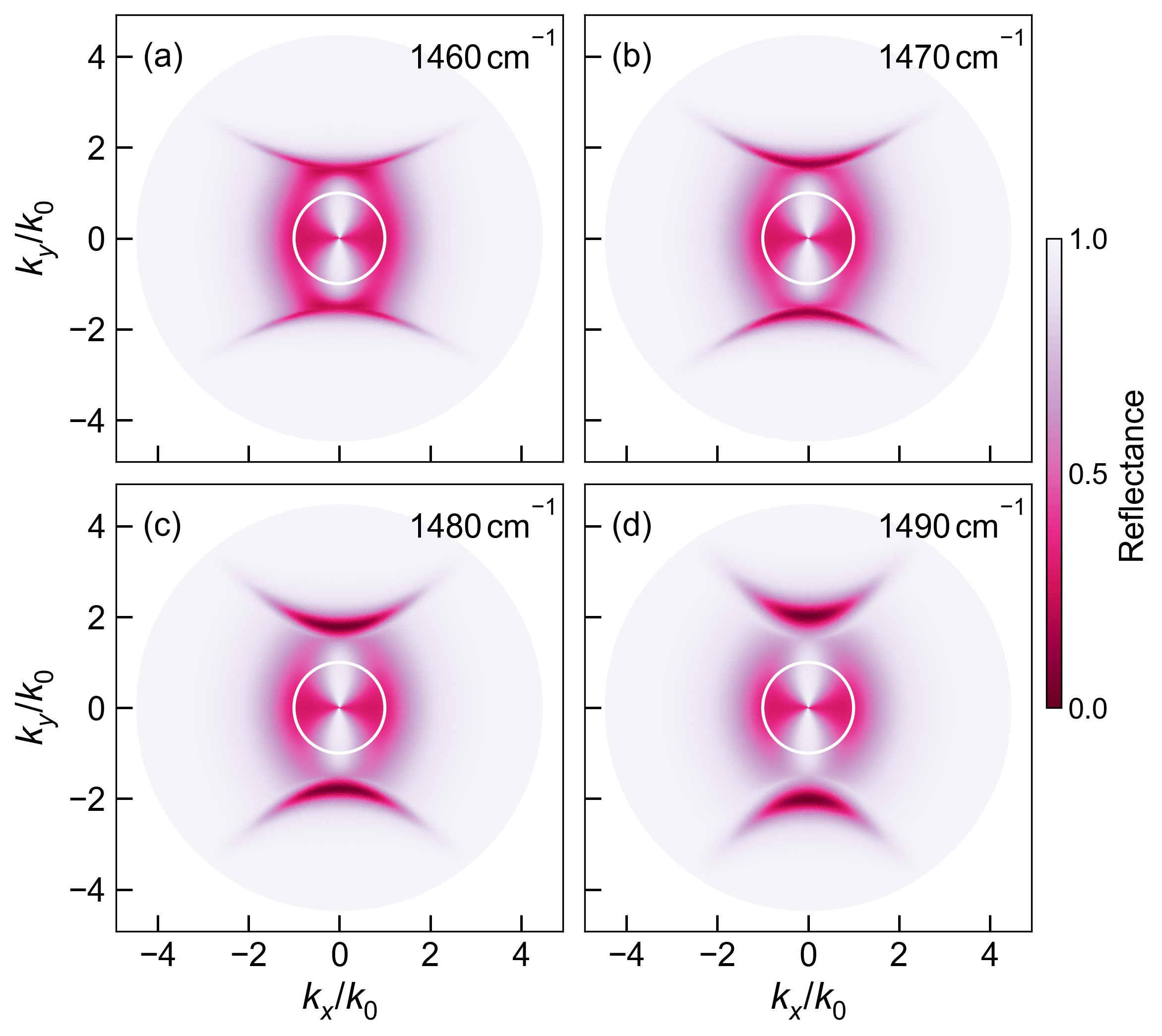}
\caption{
The frequency dependent ATR response of Calcite at with a constant air gap $d =$~0.5~$\mu$m and the anisotropy axis aligned with the 
surface ($\varphi=$~90$^{\circ}$), with the radius corresponding to $k_x$ 
where $\epsilon_p =$~18, and the azimuthal angle corresponding to the angle $\beta$.
The white circle denotes where $k_x/k_0=$~1~.
In a) $\frac{\omega}{2\pi c} =$ 1460~cm\textsuperscript{-1}.
In b) $\frac{\omega}{2\pi c} =$ 1470~cm\textsuperscript{-1}.
In c), $\frac{\omega}{2\pi c} =$ 1480~cm\textsuperscript{-1}.
In d), $\frac{\omega}{2\pi c} =$ 1490~cm\textsuperscript{-1}.
}
\label{fig:CalciteGhostFrequency}
\end{figure}

For validation of our data, we turn to calcite, where the GHP was originally observed \cite{ma_ghost_2021}.
As mentioned previously, the positive $\varepsilon_{\parallel}$ remains at ~2.3 throughout the Type II hyperbolic region.
We will investigate the range $\frac{\omega}{2\pi c} =$ 1460~cm\textsuperscript{-1} (where $\varepsilon_{\perp} =$ ~-4.8)
to 1490~cm\textsuperscript{-1} (where $\varepsilon_{\perp} =$ ~-2).
Analysing the GHP in calcite is helpful in isolating the role of $\varepsilon_{\perp}$ in its reflectance spectra.
As these magnitudes are much lower than quartz, less in-plane momentum is required to excite the GHP. 
The results are shown in Fig.~\ref{fig:CalciteGhostFrequency}, with a constant air gap $d =$~0.5~$\mu$m and the anisotropy axis aligned along the surface ($\varphi=$~90$^{\circ}$).
As the frequency increases from 1460~cm\textsuperscript{-1} in (a) to 1490~cm\textsuperscript{-1} in (d), we can see the same behaviour is observed as in quartz, with a drop in reflectivity for four emission lobes..
The 'closing' of these branches mirrors the open angle $\alpha$ in the surface-propagating rays of the original GHP discovery \cite{ma_ghost_2021}.

\section*{Appendix G: Cross Polarisation Conversion of the Leaky Hyperbolic Polariton}

\renewcommand{\theequation}{G\arabic{equation}}
\renewcommand{\thetable}{G\arabic{table}}
\renewcommand{\thefigure}{G\arabic{figure}}
\setcounter{equation}{0}
\setcounter{table}{0}
\setcounter{figure}{0}
\setcounter{page}{0}

\begin{figure*}[!ht]
\centering
\includegraphics[width=\linewidth]{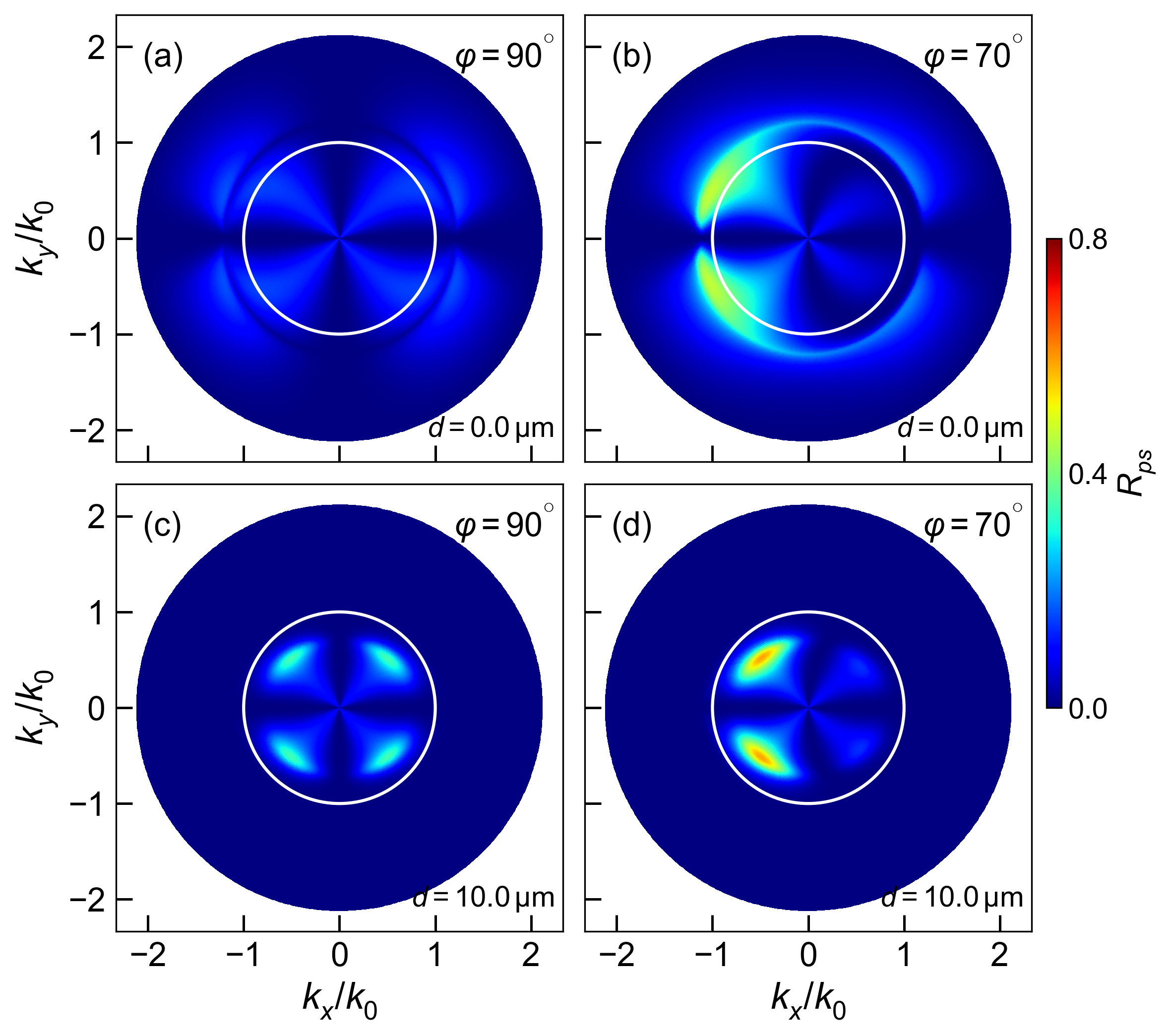}
\caption{
The cross-polarisation conversion ($R_{ps}$) induced by the LHP in Quartz at $\frac{\omega}{2\pi c} = 545\, \text{cm}^{-1}$. This quantity is the amount of s-polarised light reflected off the surface from incident p-polarised radiation.
The radius corresponding to $k_x$ where $\epsilon_p =$~4.5, and the azimuthal angle corresponding to the angle $\beta$.
The white circle denotes where $k_x/k_0=$~1~.
In a) there is no air gap ($d =$~0~$\mu$m) and $\varphi=$~90$^{\circ}$.
In b) the anisotropy orientation is unchanged ($\varphi=$~90$^{\circ}$) and an air gap is introduced to study the LHP, with $d =$~10~$\mu$m.
In c), the air gap is removed ($d =$~0~$\mu$m) and the anisotropy is rotated to $\varphi=$~70$^{\circ}$ to alter the hyperbolic dispersion.
In d), the anisotropy orientation is unchanged ($\varphi=$~70$^{\circ}$) and an air gap is introduced again to study the LHP, with $d =$~10~$\mu$m.
}
\label{fig:Leakycrosspolarisation}
\end{figure*}

As mentioned previously, asymmetric cross polarisation conversion has previously been observed in bulk Type I hyperbolic dispersion \cite{wu_asymmetric_2022}. 
In Fig.~\ref{fig:Leakycrosspolarisation}, we show the value $R_{ps}$, the proportion of the reflected radiation that possesses s-polarisation from incident p-polarised radiation, associated with Type I hyperbolic dispersion and the LHP.

In Fig.~\ref{fig:ghostcrosspolarisation}(a),where there is no air gap ($d =$~0~$\mu$m) and $\varphi=$~90$^{\circ}$, cross polarisation conversion is minimal, totally symmetric and extends outside of the free space light cone (denoted by the white circle).

When we tilt the anisotropy to $\varphi=$~70$^{\circ}$ in Fig.~\ref{fig:ghostcrosspolarisation}(b), asymmetry is introduced in the cross polarisation conversion, which is identical to the process previously observed in \cite{wu_asymmetric_2022}.
By introducing an air gap of $d =$~10.~$\mu$m [see Fig.~\ref{fig:ghostcrosspolarisation}(c) where $\varphi=$~90$^{\circ}$] we can see how the cross-polarisation conversion has increased to a value of around ~0.35, and is totally symmetric. 
Interestingly, this is within the free-space light cone, showing how evanescent behaviour from introducing the air gap is influencing polarisation, despite this being at incident angles smaller than what would induce total internal reflection.
Tilting the anisotropy to $\varphi=$~60$^{\circ}$ in Fig.~\ref{fig:ghostcrosspolarisation}(d), $R_{ps}$ becomes much more noticeably asymmetric, reaching a maximum of around ~0.7 for negative $k_x$ values, and a maximum of around ~0.1 for positive $k_x$ values.

\section*{Appendix H: Frequency Dependence of the Leaky Hyperbolic Polariton}

\renewcommand{\theequation}{H\arabic{equation}}
\renewcommand{\thetable}{H\arabic{table}}
\renewcommand{\thefigure}{H\arabic{figure}}
\setcounter{equation}{0}
\setcounter{table}{0}
\setcounter{figure}{0}
\setcounter{page}{0}
\begin{figure}[!ht]
\centering
\includegraphics[width = \linewidth, keepaspectratio]{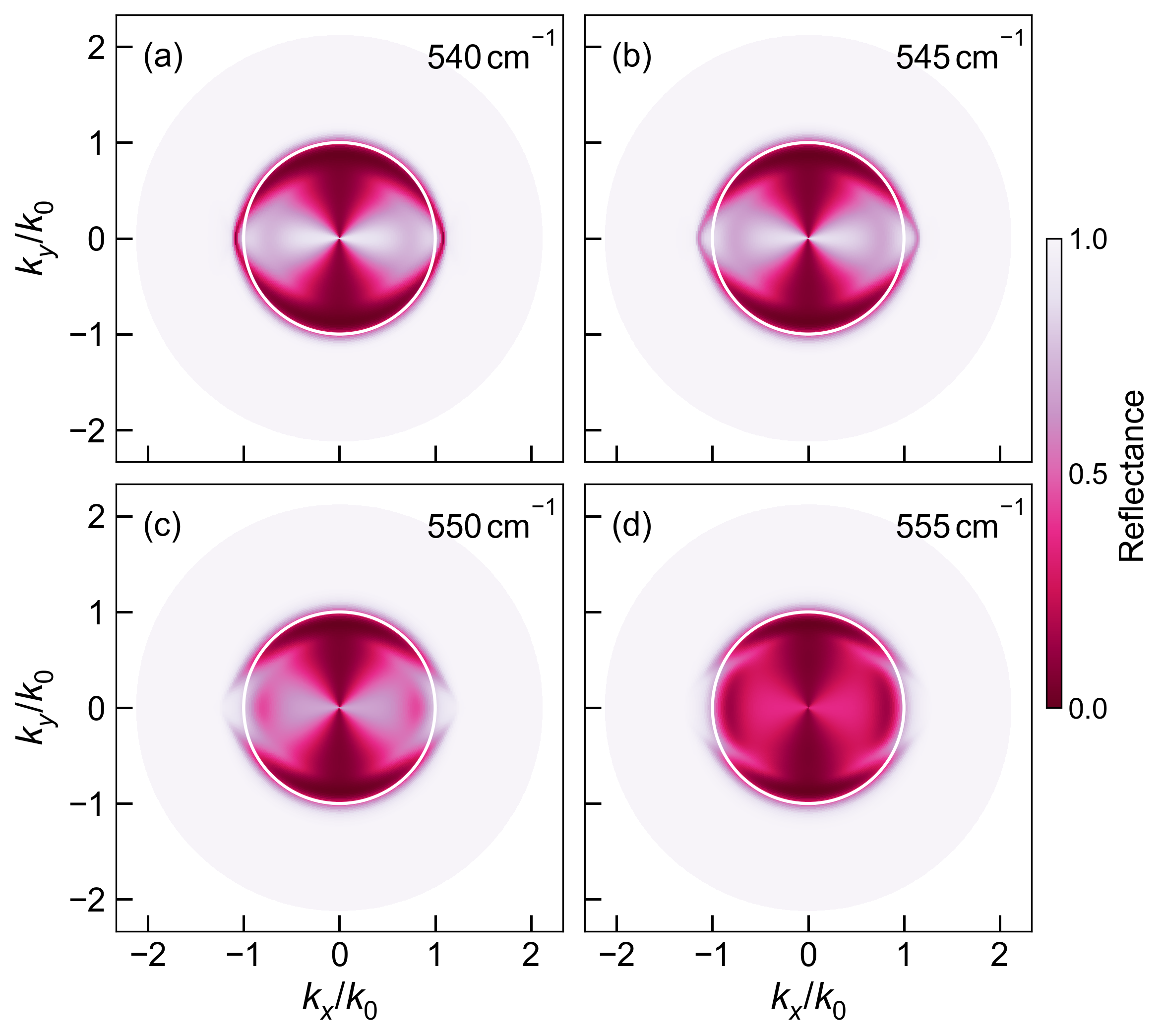}
\caption{
The frequency dependent ATR response of Quartz at with a constant air gap $d =$~10.~$\mu$m and the anisotropy axis aligned with the 
surface ($\varphi=$~90$^{\circ}$), with the radius corresponding to $k_x$ 
where $\epsilon_p =$~4.5, and the azimuthal angle corresponding to the angle $\beta$.
The white circle denotes where $k_x/k_0=$~1~.
In a) $\omega/2\pi c=$~540~cm\textsuperscript{-1}.
In b) $\omega/2\pi c=$~545~cm\textsuperscript{-1}.
In c), $\omega/2\pi c=$~550~cm\textsuperscript{-1}.
In d), $\omega/2\pi c=$~555~cm\textsuperscript{-1}.
}
\label{fig:QuartzLeakyFrequency}
\end{figure}

We have shown how LHPs can be supported in the Type I hyperbolic region and ENZ region in Quartz.
With a constant air gap $d =$~10~$\mu$m and the anisotropy axis aligned with the surface ($\varphi=$~90$^{\circ}$), in Fig.~\ref{fig:QuartzLeakyFrequency}, we show how the leaky polariton is supported at different frequencies in this region, from ~540~cm\textsuperscript{-1} to ~555~cm\textsuperscript{-1}. 
As the frequency increases within this range, $\varepsilon_{\parallel}$ increases from ~-0.64 to positive ~0.26, while the positive $\varepsilon_{\perp}$ increases from ~1.33 to ~1.69.
The reflectivity begins with a closed lenticular shape, with the LHP supported very closely to the light line (the white circle). As the frequency increases, the points of this lenticular shape move outward to higher $k_x$ values and begin to open as $\varepsilon_{\parallel}$ becomes positive, moving further apart.
This shows how in the ENZ region, the leaky polariton propagates at an angle in the $x$-$y$ plane. When $\varepsilon_{\parallel}$ is negative, this canalises the polariton along the $x$ axis.

\section{References}
\bibliography{references}
\end{document}